\documentclass[aps,pre,twocolumn,groupedaddress,showpacs,showkeys,amsmath,amssymb,flushbottom]{revtex4-1}
\usepackage{graphicx}
\usepackage{pstricks-add, pst-text}
\usepackage[utf8]{inputenc}
\usepackage{MnSymbol}

\begin{document}

\newcommand{\bskipdm}{\mskip -3.8\thickmuskip}
\newcommand{\bskiptm}{\mskip -3.0\thickmuskip}
\newcommand{\bskipsm}{\mskip -3.1\thickmuskip}
\newcommand{\bskipssm}{\mskip -3.1\thickmuskip}
\newcommand{\fskipdm}{\mskip 3.8\thickmuskip}
\newcommand{\fskiptm}{\mskip 3.0\thickmuskip}
\newcommand{\fskipsm}{\mskip 3.1\thickmuskip}
\newcommand{\fskipssm}{\mskip 3.1\thickmuskip}
\newcommand{\pint}{\mathop{\mathchoice{-\bskipdm\int}{-\bskiptm\int}{-\bskipsm\int}{-\bskipssm\int}}}
\newcommand{\ds}{\displaystyle}

\title{Selection theory of free dendritic growth in a potential flow }

\author{Martin von Kurnatowski${}^{1}$, Thomas Grillenbeck${}^{2,3}$, and Klaus Kassner${}^{1}$} 
\affiliation{${}^{1}$Institut für Theoretische Physik, \\
  Otto-von-Guericke-Universität
  Magdeburg, Germany} \email{Klaus.Kassner@ovgu.de}

\affiliation{${}^{2}$Hochschule für angewandte Wissenschaften -- Fachhochschule Rosenheim,\\
Fakultät für Angewandte Natur- und Geisteswissenschaften,\\
Rosenheim, Germany}

\affiliation{${}^{3}$Ignaz-Günther-Gymnasium, Rosenheim, Germany}

\begin{abstract}
  The Kruskal-Segur approach to selection theory in diffusion-limited
  or Laplacian growth is extended via combination with the Zauderer
  decomposition scheme. This way nonlinear bulk equations become
  tractable. To demonstrate the method, we apply it to two-dimensional
  crystal growth in a potential flow. We omit the simplifying
  approximations used in a preliminary calculation for the same system
  [T. Fischaleck, K.  Kassner, \emph{EPL} \textbf{81}, 54004 (2008)],
  thus exhibiting the capability of the method to extend
  mathematical rigor to more complex problems than hitherto
  accessible.
\end{abstract}
\date{\today}   

 \pacs{ {47.54.-r}; 
   {81.10.Aj}; 
   {11.10.Jj} 
 } 

\maketitle
\allowdisplaybreaks

\newcommand{\abl}[3][\empty]{\frac{\mathrm{d}^{#1}#2}{\mathrm{d}{#3}^{#1}}} 
\newcommand{\pabl}[3][\empty]{\frac{\partial^{#1}#2}{\partial{#3}^{#1}}} 
\newcommand{\nablaop}{\nabla} 
\newcommand{\EXP}[1]{\mathrm{e}^{#1}} 
\newcommand{\D}{\mathrm{d}} 
\newcommand{\I}{\mathrm{i}} 
\newcommand{\lr}[1]{\left(#1\right)} 
\newcommand{\lre}[1]{\left[#1\right]} 
\newcommand{\abs}[1]{\left\lvert #1 \right\rvert}  
\renewcommand{\vec}[1]{\mathbf{#1}}
\newcommand{\laplace}{\nabla^2}
\newcommand{\vecw}{\vec W}
\newcommand{\vecv}{\vec V}
\newcommand{\vecr}{\vec r}

\section{Introduction}

Pattern formation is ubiquitous in nature. Snowflakes constitute an
everyday-paradigm of a self-organized structure, apparently the first
that was the subject of scientific study \cite{kepler1611}.
Any physical pattern possesses at least one characteristic length
scale, and if it is dynamic, it also has a characteristic time scale.
The foremost task of scientific endeavour in the field of pattern
formation is to explain the emergence of these scales and to determine
them quantitatively. Since systems with linear dynamics will, due to
the superposition principle, not normally single out a particular
length scale, an essential ingredient of pattern-forming systems is
the nonlinearity of their dynamics \footnote{Linear systems may display
interesting patterns due to boundary conditions. Chladni figures are a
well-known example.  However, we rather speak of pattern formation, when
scale selection is intrinsic to the dynamics.}.

As it turns out, snowflake-like structures -- dendritic morphologies
-- also arise at microscopic scales in the casting of metals and they
determine structural properties such as the strength of the material,
which imparted considerably more importance to scientific
preoccupation with them than just fundamental interest would have.

The first models of dendritic crystal growth assumed transport of heat
away from or material to, the growing nucleus to be simply diffusive,
so the term diffusion-limited growth was coined. Understanding the
selection of dynamical features such as a basic length scale and the
growth velocity turned out to be remarkably difficult even within
these simplifying models. Almost forty years passed between Ivantsov's
approximate solution \cite{ivantsov47} that did not exhibit selection
and the development of an analytic theory explaining the mechanism of
structure selection \cite{langer86b,caroli86,benamar86}.  This may
seem even more surprising considering that the bulk equations of
diffusion-limited systems are \emph{linear} and the nonlinearity of
the dynamics emerges solely via the equations of motion for the
two-phase interface. In fact, the analytic approaches developed had
to rely heavily on this linearity.

Ivantsov's theory, neglecting surface tension at the boundary between
the melt and the solid, predicts only the product of the tip radius of
a dendrite and its growth velocity, as a function of the undercooling.
In experiments, the undercooling determines both quantities
separately.  A decisive step towards the solution was the insight that
without surface tension the problem is ill-posed
\cite{benjacob84b,kessler85} and that the capillary length has to be
taken into account, even if it is much smaller than any length scale
of the arising pattern. Surface tension regularizes the mathematical
problem and drastically alters the solution space. Without surface
tension, there is a continuum of parabolic needle crystal solutions.
With \emph{isotropic} surface tension, there are \emph{no solutions}
with a shape close to one of these Ivantsov parabolas (or
paraboloids), no matter how small the surface tension, a fact that
testifies to the singular nature of the ``perturbation'' surface
tension. With \emph{anisotropic} surface tension, the continuum of
Ivantsov solutions is reduced to a discrete set, with the fastest of
the needle crystal solutions being the only linearly stable one
\footnote{What is said here for surface tension, holds, \emph{mutatis
    mutandis}, also for interfacial kinetics. With an anisotropic term
  for the velocitity-dependent deviation of the interface temperature
  from its equilibrium value, selection happens even if the
  Gibbs-Thomson effect is not taken into account
  \protect\cite{brener91}. If both surface tension and the kinetic
  term are isotropic, there is no selection of parabolic shapes in
  free growth.}. Hence, the selection problem is broken down into two
parts -- an existence problem for a discrete set of solutions and the
stability analysis singling out one element of the set as the one that
should be observed. A completely analogous theory was developed for
Saffman-Taylor fingers in viscous fingering
\cite{hong86,shraiman86,combescot86}, where selection is also due to
surface tension, albeit not, of course, to its anisotropy.

These theories were two-dimensional, just as the original
numerical work giving evidence for a selection mechanism based on
solvability \cite{meiron86a,kessler86,kessler86d}. Three-dimensional
situations considered initially referred to axisymmetric crystals
\cite{kessler86,barbieri87}, hence were not very realistic. Later,
steps were taken to extend the theory towards non-axisymmetric
needle-crystal shapes \cite{kessler87,kessler88a} and eventually, an
analytic theory was developed for the fully non-axisymmetric case
\cite{benamar93,brener93a}; all the necessary elements of the final
complete theory were not present before Ref.~\onlinecite{brener93a}.

From the outset, two different analytic approaches were pursued. With
the first method, the equation of motion of the two-phase interface is
linearized, which leads to an integro-differential equation in
non-local problems (such as dendritic growth or viscous fingering).
Using Fredholm's alternative, a solvability condition is derived that
is satisfied only by a discrete set of values of the selection
parameter (a nondimensionalized surface tension or its inverse)
\cite{langer86,hong86,barbieri87}. The second approach, pioneered by
Kruskal and Segur \cite{kruskal91}, consists in solving the interface
equation far from singular points in the complex plane via a
perturbation expansion in terms of the small selection parameter and
in the vicinity of these points via a scale transformation and
reduction to a local equation. The two solutions then have to be
asymptotically matched to obtain a globally valid solution. Parameter
relationships established in the matching procedure yield the
selection criterion
\cite{benamar86,tanveer89,benamar90,brener91,tanveer00}. There is
general agreement that only the second approach is mathematically
rigorous \cite{combescot86,hong87,barbieri89,tanveer89}. The linearization of the
first method introduces approximations that normally will not
invalidate the scaling relations obtained; but it will not reproduce
their prefactors correctly nor provide a globally valid approximate
solution. Moreover, if the equations contain more than one small
parameter (say, a kinetic coefficient or a characteristic number
describing the flow, besides the usual selection parameter), the
linearization may produce even worse results due to the
structural instability of the problem \cite{tanveer89}.

In both approaches, it is necessary to first derive an
(integro-differential) equation for the interface position depending
on a single independent variable. This can be achieved, e.g., by
eliminating the bulk field variables via conformal mapping (in the
viscous fingering case) or using Green's function methods (in crystal
growth). These techniques are applicable only for linear bulk
equations, which seemed to preclude utilization of the method for
convection problems.

For a long time, the only exception to this restriction has been the
work on two-dimensional crystal growth in an Oseen flow by Bouissou
and Pelcé \cite{bouissou89b}. To obtain the selection criterion, they
used a method, outlined in Ref.~\onlinecite{pelce88}, which is closely
related to the first of the two approaches mentioned, hence not
rigorous.  Their method has recently been extended by Alexandrov et
al.~\cite{alexandrov10} to include solute diffusion. The equation of
motion for the deviation of the solution from that of the problem
without surface tension is simplified in the style of a linear
stability analysis which allows to avoid the derivation of an
integro-differential equation. Moreover, the adjoint linear operator
is constructed heuristically in Fourier space to obtain a solvability
condition in the spirit of the Fredholm alternative, a procedure that
may introduce additional (possibly problematic) approximations.

A different method, having the potential of achieving the same level
of rigor for problems with nonlinear bulk equations as the asymptotic
matching approach, was recently introduced \cite{fischaleck08,fischaleck08b}. It
consists in a combination of Zauderer's decomposition scheme
\cite{zauderer78} for partial differential equations with the
Kruskal-Segur approach. Zauderer decomposition is the step allowing
reduction of nonlinear bulk equations to an interface equation and
thus circumventing the necessity of an exact integral equation, available
only for linear bulk equations.  Reference~\onlinecite{fischaleck08},
dealing with potential flow, was more or less a proof of concept, in
which we copiously used additional approximations to simplify the
result to an easily digestable form, allowing to map it in the end to
the flowless finite Péclet number case treated by Ben Amar
\cite{benamar90}. The main purpose of the present paper is to remove
these approximations, which renders the treatment more complex, but
does not impose insurmountable obstacles. Of course, the mapping
obtained gets lost, because it was only approximate. Clearly,
potential flow is not a very realistic assumption, but it has the
advantage that the unperturbed problem is exactly solvable; the analog
of Ivantsov's analytic solution exists. This is different in the case
treated by Bouissou and Pelcé \cite{bouissou89b}, where already the
zeroth-order problem is solved within an approximation, replacing the
Navier-Stokes equations with the Oseen problem (which in two
dimensions does not even give a uniform approximation to the true
flow \cite{saville88}). We will consider more realistic flows and a better approach than
the Oseen approximation in a later publication.

The paper is organized as follows. In Sec.~\ref{sec:modeleqs}, the
model equations are given. They are nondimensionalized and rewritten
in parabolic coordinates in Sec.~\ref{sec:parabolic_coord}.
Section~\ref{sec:ivantsov_like_sol} gives the analog of the Ivantsov
solution in the presence of a potential flow. Then the method of
Zauderer decomposition is explained in Sec.~\ref{sec:zauderer_decomp},
allowing us to reduce the set of partial differential equations of the
full problem to an integro-differential equation for the interface
alone, without losing the terms decisive for solvability theory.
Next, the decomposed equations are solved to first order (in the small
parameter $\sigma^{2/7}$) in Sec.~\ref{sec:sol_decompos}. Near the
solid-liquid interface, the relevant behavior beyond all orders of
regular perturbation theory is obtained in Sec.~\ref{sec:wkb} from a
WKB analysis.  On the other hand, in Sec.~\ref{sec:sol_near_sing}, the
asymptotic Kruskal-Segur reduction to a locally valid equation,
applicable near a singularity in the complex plane, is carried out. It
leads to a nonlinear integro-differential equation constituting an
eigenvalue problem.  The numerical solution of this eigenvalue problem
determines the selected velocity (and other properties) of the needle
crystal.  Detailed results are given for a set of parameters
corresponding to a particular experimental system, which however does
not exhibit potential flow, so the comparison is only qualitative.
Some conclusions are offered in Sec.~\ref{sec:conclusions}. A few
general calculations and slightly elaborate mathematical conversions
are relegated to two appendices.

\section{Model equations}
\label{sec:modeleqs}

Heat transport in the liquid and solid phases is described by the
diffusion-advection equations\\
\begin{equation}
 \pabl{T}{t}+\left(\vec w\cdot\nablaop\right)T=D\laplace T
\label{eq:diffadvect}
\end{equation}
with $\vec w$ the flow velocity in the liquid and $\vec w\equiv 0$ in
the solid. The advection term coupling the temperature and flow
equations renders the bulk problem nonlinear, despite the simplifying
assumption of potential flow made below (which reduces the flow
description to a linear equation).  As the notation suggests, we assume
the thermal diffusivity $D$ to be the same in both phases (symmetric
model).  We consider an
incompressible flow, which means that a stream function can be
introduced. In two dimensions, its defining equation takes the form
\begin{align}
 \vec w=\nablaop\times(\psi\vec e_z) = \psi_y\vec e_x-\psi_x\vec e_y \>. 
\label{eq:def_stream_function}
\end{align}
Because $\psi$ depends on $x$ and $y$ only, we obtain
$\nablaop\times\vec w = -\laplace \psi\,\vec e_z $ and taking the flow to be
potential, we have
\begin{align}
\laplace \psi =0 \>. 
\label{eq:stream_function}
\end{align}

When specializing Eq.~\eqref{eq:diffadvect} to one of the phases, we
will denote the temperature variable by $T^l$ and $T^s$ in the liquid
and the solid, respectively.  Equations \eqref{eq:def_stream_function}
and \eqref{eq:stream_function} are the bulk equations of motion for
the flow.

Because we are looking for steady-state solutions, we need not
prescribe detailed initial conditions. We must however specify
boundary conditions for each of the bulk equations. At infinity in
either the liquid or solid we require homogeneous Dirichlet boundary
conditions for the temperature fields, i.e., we set the temperature
constant:
\begin{subequations}
\begin{align}
T^l(\vec x) &\to T_\infty  &&\text{for } d(\vec x,\Gamma) \to \infty\>,
\label{eq:templiq_bc_inf}\\
T^s(\vec x) &\to T_M  &&\text{for } d(\vec x,\Gamma) \to -\infty\>.
\label{eq:tempsol_bc_inf}
\end{align}
\end{subequations}
Herein, $\Gamma$ is the interface and $d(\vec x,\Gamma)$ denotes a
signed distance function, increasing towards the liquid. $T_M$ is the
bulk melting temperature of the solid, and for crystal growth to occur
in a pure system, we must have $T_\infty< T_M$. The dimensionless
parameter characterizing this undercooling is
\begin{align}
\Delta = \frac{T_M- T_\infty}{L/c_p}\>,
\end{align}
where $L$ and $c_p$ are the latent heat and specific heat, both
referred to a unit volume.

Moreover, there are boundary conditions at the interface, reading
\begin{subequations}
\begin{align}
 &T^s\vert_\Gamma=T^l\vert_\Gamma\>, \label{eq:temp_continuous_dim}\\
 &T^l\vert_\Gamma=T_M-\frac{L}{c_p}d_0\kappa a(\theta)\>,
\label{eq:Gibbsstart}\\
 &L V_n=D c_p\left[\nablaop T^s\vert_\Gamma-\nablaop T^l\vert_\Gamma\right]\cdot\vec n
 \label{eq:Stefanstart}\>.
\end{align}
\end{subequations}
Equation \eqref{eq:temp_continuous_dim} describes continuity of the
temperature across the interface, Eq.~\eqref{eq:Gibbsstart} is the
Gibbs-Thomson condition giving the equilibrium temperature of a
melt-crystal interface with curvature $\kappa$. That is, we assume
kinetic effects to be negligible, implying local thermal equilibrium
at the interface.  If the interface is given by $y=h(x)$, then $\kappa
= -h''(x)/\left(1+h'(x)^2\right)^{3/2}$.  $d_0$ is the average
capillary length,
\begin{align}
d_0 = \gamma_0 \frac{T_M c_p}{L^2}\>,
\label{eq.capillary_length}
\end{align}
where $\gamma_0$ is the angular average of the orientation dependent
surface tension $\gamma(\theta)$. $\theta$ is the angle of the
interface normal with some fixed direction, for example the direction
of the $y$ axis. From the thermodynamics of interfaces we know
\cite{nozieres92} that the angular dependence of the non-averaged
capillary length is not that of the surface tension but that of the
surface stiffness $\gamma(\theta)+\gamma''(\theta)$. This is described
by the factor $a(\theta) \equiv
(\gamma(\theta)+\gamma''(\theta))/\gamma_0$. For simplicity, we will
assume fourfold anisotropy here, described by a single harmonic, i.e.
\begin{equation}
  a(\theta)= 1-\beta \cos(4\theta)\>.
\label{eq:def_aniso}
\end{equation}

From a mathematical point of view,
Eqs.~(\ref{eq:temp_continuous_dim},\ref{eq:Gibbsstart}) together with the
boundary conditions at infinity and some initial condition for the two
temperature fields are sufficient to solve the diffusion problem in
the two phases, with given flow field and interface position.
However, the interface position is a priori unknown, its determination
is part of the problem. Therefore, an additional interface equation is
needed. This is the third boundary condition, the Stefan
condition \eqref{eq:Stefanstart}.  Physically, it follows from energy
conservation across the interface. $V_n$ is the interface normal
velocity and the left-hand term describes latent heat production due
to advancement of the interface, whereas the right-hand side gives the
sum of the heat currents into the solid and liquid phases. The normal
vector $\vec n$ points from the solid into the liquid.

The flow field $\vec w$ is dynamic only in the liquid. We need a
boundary condition at infinity, where we impose a constant flow directed
opposite to the growth direction of the needle crystal:
\begin{align}
\vec w(\vec x) &\to -U \vec e_y &&\text{for } d(\vec x,\Gamma) \to \infty\>.
\label{eq:flow_bc_inf}
\end{align}
Because we have potential flow, we cannot impose conditions for all
three components of the flow velocity at the interface; the Laplace
equation for the stream function does not admit prescription of more
than one scalar quantity. Physically, this is reasonable, since
potential flow is frictionless, hence we cannot prescribe the
tangential velocity at the interface, there is no \emph{no-slip}
condition. The normal velocity, on the other hand, follows from mass
conservation. We assume the simplest case, \emph{viz.}~equal mass
densities in the solid and the liquid. Then the liquid is neither
sucked towards the solid (which would be the case if the density of
the solid were higher than that of the liquid), nor ejected from it.
Since the solid does not move \footnote{Meaning that no volume element
  of the solid is in motion. The \emph{interface} moves, of course,
  due to the addition of solid.}, the normal velocity of the liquid
must be zero:
\begin{align}
\vec n\cdot\vec w(\vec x)\vert_\Gamma   = 0\>.
\label{eq:flow_bc_int}
\end{align}

Equations \eqref{eq:diffadvect} through \eqref{eq:flow_bc_int},
supplemented by initial conditions for all the fields, constitute the
complete mathematical description of an idealized physical system.
Requiring the solution to be stationary and to correspond to a crystal
growing at constant velocity $V$ along the $y$ direction we may
replace $V_n$ in Eq.~\eqref{eq:Stefanstart} by $V \vec e_y \cdot\vec n = V
n_y$. Transforming to a moving frame of reference,
\begin{subequations}\begin{align}
 \vec r&\to\vec r +Vt\vec e_y\>,\\
 \vec w&\to\vec w + V\vec e_y\>,
\end{align}
\end{subequations}
in which the interface is at rest, all time derivatives $\partial_t$
get replaced by $-V \partial_y$. Note that due to the transformation
of $\vec w$, Eq.~\eqref{eq:diffadvect} is invariant under this change
of frame. Nevertheless, the time derivative can be dropped after the
transformation, because we seek a time-independent solution. Moreover,
the ``flow velocity in the solid'' becomes $-V \vec e_y$ by virtue of the
transformation.

\section{Parabolic coordinates and nondimensionalization}
\label{sec:parabolic_coord}

A family of exact analytic steady-state solutions to the model
equations exists for vanishing capillary length and, similar to
Ivantsov's solution in the flowless case, the crystal interface is
parabolic \cite{das84,benamar88c,cummings99}. Therefore, it is useful
to introduce parabolic coordinates. We employ \emph{conformal}
parabolic coordinates
\begin{equation}
x=\eta \xi \>, \qquad y = \frac12 \left(\eta^2-\xi^2\right)\>,
\end{equation}
their advantage being equality of the scale factors
$g_\xi=g_\eta=\sqrt{\xi^2+\eta^2}$. In the appendix, some of the
transformation formulas and a graphical representation of the
coordinate lines are given.

To nondimensionalize the equations, we use the tip radius $\rho$ of
the Ivantsov like solution, defined as the inverse of the curvature at
the tip, as a length scale. The corresponding diffusion time
$\rho^2/D$ is taken as a time scale
\begin{equation}
 x,y\to\rho x,\rho y,\;\kappa\to\frac{\kappa}{\rho}\,,\quad t\to\frac{\rho^2}{D}t\>.
\end{equation}
Note that this implies $\xi$ and
$\eta$ to scale with $\sqrt{\rho}\,$: $\xi,\eta\to
\sqrt{\rho}\,\xi,\sqrt{\rho}\,\eta$.  

The nondimensional form of the flow
velocity follows immediately
\begin{equation}
\vec w \to\frac{D}{\rho}\vec w \>.
\end{equation}
(This implies $\psi\to D\psi$.) The nondimensional flow velocity at
infinity then becomes
\begin{equation}
P_f \equiv \frac{\rho U}{D} \>,
\label{eq:def_flowpeclet}
\end{equation}
the so-called \emph{flow Péclet number}. Moreover, it will turn out useful to
include the \emph{growth Péclet number}
\begin{equation}
P_c \equiv \frac{\rho V}{D} \>,
\label{eq:def_growpeclet}
\end{equation}
into the prescription for nondimensionalization of temperature
\begin{equation}
 T\to T_M+\frac{L}{c_p} P_c T\>,
\end{equation}
which means that the nondimensional temperature in the liquid
approaches $-\Delta/P_c$ at infinity.

With these transformations and dropping time derivatives, as we are
interested in stationary solutions, we obtain the following set of bulk equations
\begin{subequations}\begin{align}
 \psi_\eta T^l_\xi-\psi_\xi T^l_\eta&=T^l_{\xi\xi}+T^l_{\eta\eta} \,,
\label{eq:diffliq2}\\
P_c\left(\xi T^s_\xi-\eta T^s_\eta\right)&=T^s_{\xi\xi}+T^s_{\eta\eta} \,,
\label{eq:diffsol2}\\
\psi_{\xi\xi}+\psi_{\eta\eta}&=0\>.
\label{eq:stream2}
\end{align}
\end{subequations}
 The boundary conditions for the  fields at the interface now read
\begin{subequations}
\begin{align}
  T^s&=T^l\>,
 \label{eq:temperatur_cont_parab}
  \\
  T^l&=-\frac 12\sigma\kappa a(\theta)\>,
 \label{eq:Gibbs_Thomson_parab}
  \\
  \left[\xi\eta_s\right]'&=-\eta_s'\left(T_\xi^s-T_\xi^l\right)+T_\eta^s-T_\eta^l\>,
  \label{eq:Stefan_parab}
  \\
  \psi_\xi+\eta_s'\psi_\eta&=P_c\left(\eta_s+\eta_s'\xi\right)\>,
  \label{eq:masscon05}
\end{align}
\end{subequations}
where $\eta_s(\xi)$ is the interface position, a  prime
means a derivative with respect to $\xi$ along the interface, and
\begin{equation}
\sigma=\frac{2d_0}{\rho P_c}
\label{eq:def_sigma}
\end{equation}
is the selection or stability parameter. (In order not to overburden
the notation, we have dropped the qualifier $\vert_\Gamma$ next to the
fields and their derivatives, indicating that these quantities have to
be evaluated at the interface position.)

Finally, the boundary conditions at infinity may be written (see
Fig.~\ref{fig:coordinates})
\begin{subequations}
\label{eq:bc_inf2}
\begin{align}
  T^l&\to - \frac{\Delta}{P_c} && (\eta \to \infty)\>,
\label{eq:bc_tliq_inf}
  \\
  T^s&\to 0 &&  (\eta<1,\> \abs{\xi} \to \infty)\>,
\label{eq:bc_tsol_inf}
  \\
  \psi &\sim \left(P_f+P_c\right)\eta\xi  &&(\eta \to \infty)\>,
\label{eq:bc_psi_inf}
\end{align}
\end{subequations}
the last equation being an asymptotic equality \cite{bender78}.

\section{Exact solution in the absence of surface tension}
\label{sec:ivantsov_like_sol}

If we neglect surface tension in the spirit of Ivantsov,
eqs.~(\ref{eq:temperatur_cont_parab},\ref{eq:Gibbs_Thomson_parab})
tell us that with vanishing capillary length the interface becomes an isotherm,
$T^l\vert_\Gamma=T^s\vert_\Gamma=0$. Assuming it to be a
coordinate line suggests the temperature field to depend on one of the
coordinates only, i.e. $T=T(\eta)$. Inserting this into
(\ref{eq:diffliq2},\ref{eq:diffsol2}), we have
\begin{subequations}
\begin{align}
 -\psi_\xi T^l_\eta &=T^l_{\eta\eta}\label{eq:potTIv1}\>,\\
 -P_c \eta T^s_\eta &=T^s_{\eta\eta}\label{eq:potTIvsol}\>.
\end{align}
\end{subequations}
In view of boundary condition \eqref{eq:bc_tsol_inf}, we see that the
second equation is solved by $T^s\equiv 0$. The first can have a
purely $\eta$ dependent solution only if
\begin{align}
\psi=\xi f(\eta)\>.
\end{align}
Inserting this into \eqref{eq:stream2}, we find 
\begin{align}
f''(\eta)=0\quad\Rightarrow\quad f(\eta)=c_1\eta+c_2\>,
\end{align}
with constants $c_1$ and $c_2$ to be determined from the two boundary
conditions (\ref{eq:masscon05},\ref{eq:bc_psi_inf}). By assumption,
the interface is at $\eta_s(\xi)\equiv 1$, so \eqref{eq:masscon05}
simplifies into $\psi_\xi(1) = P_c$. We find $c_1=P_f+P_c$ and
$c_2=-P_f$. Therefore,
\begin{align}
 \psi=\psi^{\text{Iv}}=\xi\left(P_c \eta+P_f(\eta -1)\right)
\label{eq:potPIv2}.
\end{align}
This leaves us with the ordinary second-order differential equation \eqref{eq:potTIv1}
$T^l_{\eta\eta}+\left(P_c \eta+P_f(\eta -1)\right)T^l_\eta=0$ subject to the
boundary conditions $T^l(1)=0$ and \eqref{eq:bc_tliq_inf}. The solution is straightforward:
\begin{align}
T^l =  T^{\text{Iv}}(\eta)=-\EXP{\frac{P_c}{2}}\int\limits_1^\eta\EXP{-\frac{P_c}{2}\eta'^{2}
-\frac{P_f}{2}\left(\eta'-1\right)^2}\D\eta'\>,
\label{eq:potTIv2}
\end{align}
where $P_c$ is determined as a function of $\Delta$ and $P_f$ from
\begin{align}
 \frac{\Delta}{P_c}=\EXP{\frac{P_c}{2}}\int\limits_1^\infty\EXP{-\frac{P_c}{2}
\eta'^{2}-\frac{P_f}{2}\left(\eta'-1\right)^2}\D\eta'\>.\label{eq:selection1}
\end{align}
In the limit $P_f\to 0$, this becomes identical to the usual Ivantsov
relation for diffusion-limited dendritic growth, whereas for $P_c\ll
P_f$ we can evaluate the formula analytically, which yields
$\Delta=P_c \sqrt{\pi/2 P_f}$. The selection problem arises in the
same way here as in the flowless case: only $P_c$ is determined, given
the undercooling and the imposed flow, by the Ivantsov-like solution.
Given $P_c$, we can calculate the product of the growth rate $V$ and
the tip radius $\rho$ but not both quantities separately. Ivantsov's
approach eliminates the parameter $\sigma$ from the equations. As
\eqref{eq:def_sigma} tells us, we may calculate $\rho$ once we know
$\sigma$ and $P_c$. So the aim must be to include $\sigma$ into the
theory and to obtain a value for it.

For easy reference, we will call the solution \eqref{eq:potPIv2} --
\eqref{eq:selection1}  the flow-Ivantsov
solution.

\section{Zauderer decomposition and continuation to the complex plane}
\label{sec:zauderer_decomp}
The approach to be followed is singular perturbation theory about the
flow-Ivantsov solution \footnote{As we shall see later, we do not
  precisely expand about the flow-Ivantsov solution but rather about
  an approximation to it that becomes accurate in the vicinity of the
  appropriate complex-plane singularity.}. Thus, we shall consider
 small deviations from it
\begin{align}
\label{eq:field_linearization}
 T\to T^{\text{Iv}}+T\quad\text{and}\quad\psi\to\psi^{\text{Iv}}+\psi\>,
\end{align}
but we will be careful to avoid illegitimate linearizations. In
particular, we will not linearize terms containing derivatives of the
interface position.

In this first approach, we shall restrict ourselves to the limit of
small growth Péclet number, i.e., $P_c\ll 1$. As it turns out,
interesting results for the selection parameter arise only, if we also
assume $P_c\ll P_f$. In particular, this means that terms with a
factor of $P_c$ will be neglected in the exponentials of
Eqs.~\eqref{eq:potTIv2} and \eqref{eq:selection1}.

The diffusion-advection equation \eqref{eq:diffliq2} then becomes an
inhomogeneous equation for the temperature deviation, in the liquid,
from the flow-Ivantsov solution
\begin{align}
 T^l_{\xi\xi}+T^l_{\eta\eta}&-\left(\psi_\eta+\xi P_f\right)T^l_\xi
\nonumber\\
 &+\left(\psi_\xi+P_f(\eta -1)\right)T^l_\eta=\psi_\xi
\EXP{-\frac{P_f}{2}\left(\eta-1\right)^2}\>.
\label{eq:diffliq3}
\end{align}
The only approximation in this equation is that $P_c$ has been set
equal to zero. All field nonlinearities are still present.

In the solid (Eq.~\eqref{eq:diffsol2}), we just drop the terms
multiplied by $P_c$ and obtain a Laplace equation
\begin{align}
T^s_{\xi\xi}+T^s_{\eta\eta} = 0\>,\label{eq:diffsol3}
\end{align}
and Eq.~\eqref{eq:stream2}, being linear, remains formally unchanged, 
\begin{align}
\psi_{\xi\xi}+\psi_{\eta\eta}=0\>,\label{eq:stream3}
\end{align}
but the meaning of  $\psi$ is now different (it is the deviation of
the stream function from its form in the flow-Ivantsov solution).

To rewrite the boundary conditions, we set $\eta_s=1+h(\xi)$. While it
is legitimate to view $h(\xi)$ as a small quantity, this may not
be true for $h'(\xi)$ and higher derivatives. Expanding
\eqref{eq:temperatur_cont_parab} about the flow-Ivantsov solution, we
have $T^{\text{Iv}}\vert_\Gamma+ T^l\vert_\Gamma =
\partial_\eta T^{\text{Iv}}\vert_{\eta=1}\, h + T^l(\xi,1) =
T^s\vert_\Gamma =T^s(\xi,1)$. Later, we will need the derivative of
the interface temperature with respect to $\xi$. We are then not
allowed to simply take the \emph{partial} derivative of the
temperature field with respect to $\xi$, which would give $T^l_\xi
-h'(\xi) = T^s_\xi$ (because $\partial_\eta
T^{\text{Iv}}\vert_{\eta=1}=-1$).  Actually, the derivative must be
taken along the interface, so we obtain $T^l_\xi + T^l_\eta h'(\xi) -
h'(\xi) = T^s_\xi + T^s_\eta h'(\xi)$ instead. Keeping this proviso in
mind, we may use the simpler form before differentiation as the
shortest description of the appropriate boundary condition. The full
set of interface boundary conditions then reads
\begin{subequations}
\label{eq:interf_bcall}
\begin{align}
 T^s+h&=T^l\>, \label{eq:cont1}
\\
 T^s&=-\frac 12\sigma\kappa a(\theta)\>,\label{eq:Gibbs1}
\\
 \left[\xi h\right]'&=\left(\pabl{}{\eta}-h'\pabl{}{\xi}\right)\left(T^s-T^l\right)\>, 
\label{eq:Stefan2}\\
 \psi_\xi+h'\psi_\eta&=-P_f\lre{h\xi}'\>.
\label{eq:masscon1}
\end{align}
\end{subequations}
All of these are evaluated at $\eta=1$, but if they are to be
differentiated, then this has to be done before setting $\eta=1$ and
field derivatives with respect to $\eta$ will arise.

Zauderer's asymptotic decomposition \cite{zauderer78} is a projection
scheme reducing the solution of a system of partial differential
equations (PDEs) to the solution of a series of first-order equations.
These may be decoupled within a perturbative approach, if a small
parameter or slow variable is available. (Otherwise, the series of
first-order equations does not offer any simplification over the
original system of PDEs.) Originally conceived for hyperbolic
equations, the method generalizes to elliptic systems in the complex
plane.

Asymptotic decomposition seems to have largely passed into
oblivion (at least in the physics world), possibly because often
multi-scale expansions are superior to it, leading to more easily
tractable equations.  Nevertheless, for the problem considered here,
Zauderer decomposition is particularly well-suited, not losing
information about transcendentally small terms (if the ``principal
part'' \cite{zauderer78} of the set of equations is chosen
correctly).

In \cite{fischaleck08}, we discussed that in the case of purely
diffusive transport and in the limit of small growth Péclet number,
Zauderer decomposition of the transport equations is equivalent to
their factorization
\begin{align}
  \left(\partial_\xi+\I\partial_\eta\right) T^l = 0\>,\quad
  \left(\partial_\xi-\I\partial_\eta\right) T^s = 0 \>,
\label{eq:diffusive_small_peclet} 
\end{align}
and that a few simple manipulations of these partial differential
equations using the boundary conditions lead to a local equation for
the interface position $h(\xi)$,
\begin{align}
\sigma a(\theta) \kappa = (1-\I\xi) h(\xi) \>.
\label{eq:bigeq_simple}
\end{align}
This equation contains the complete information needed to compute
the mismatch function that has to be zero at the tip of the
needle crystal for selection to be possible.  Near the singularity
$\xi=-\I$ in the complex plane, Eq.~\eqref{eq:bigeq_simple}
describes the dominant behavior of the solution. 

Equations \eqref{eq:diffusive_small_peclet} are, due to their
simplicity, well-suited for a discussion of the strategy of our
approach. Typically, Zauderer decompostion produces, from the basic
partial differential equations of the problem, a leading-order or
``principal-part'' equation in each domain that is first order and
easily solvable (for example by the method of characteristics). The
complete solution will be a sum of this leading term and other
contributions that may be calculated in subsequent steps of the
perturbative scheme. In the simple case considered here for
explanatory purposes, the temperature field satisfies the Laplace
equation factorizing in the complex plane (a formal Zauderer
decomposition just reproduces this factorization). Solving the
factorized equations, one finds
\begin{align}
T(\xi,\eta) = f_1(\xi+\I(\eta-1))+f_2(\xi-\I(\eta-1))
\end{align}
with analytic functions $f_1$ and $f_2$. Inserting this solution into
the boundary conditions at $\eta=1$ and analytically continuing the
resulting equations to $w=-\I$ (where $w$ is the analytic continuation
of $\xi$), some of their terms must become singular in the limit
$h(\xi) \ll 1$ ($\sigma\to 0$), because the curvature term in
\eqref{eq:bigeq_simple} becomes singular in that limit [see
Eq.~\eqref{eq:curvature}]. The solution for $T^l$ must be analytic in the
liquid, i.e., for $\eta>1$, corresponding to the upper half complex
plane. Hence the $f_2$ term will remain analytic near $w=-\I$
\footnote{The argument $-\I$ to $f_2$ obtained by setting $w=\xi=-\I$
  and $\eta=1$ can alternatively be constructed setting $\xi=0$ and
  $\eta=2>1$.}. The important contribution to $T^l$ that may diverge
near the singularity will then come from $f_1$, and this function is
the solution to the first equation of
\eqref{eq:diffusive_small_peclet}. Similarly, it may be argued that in
the solid the $f_2$ term is the important one; it solves the second
equation of \eqref{eq:diffusive_small_peclet}. Thus,
Eqs.~\eqref{eq:diffusive_small_peclet} give us a solution that is
valid near the singularity. At the interface, far away from the
singularity, this approximation is also justified (albeit to a lesser
degree), because there the curvature term in the boundary conditions
can be linearized. For the (homogenous part of) the linearized
equations, each of the $f_1$ and $f_2$ terms alone is a solution. To
obtain the full solution, the second singularity at $w=\I$ has to be
taken into account. To treat that case, the lowest order equations for
$T^l$ and $T^s$ would have to be interchanged. So the choice of the
sign of the $\partial_\eta$ term in the factorization depends on the
singularity considered, and the signs for the liquid and solid domains
must be opposite to each other. In our simple example, it is
sufficient to just consider one singularity, because the result for
the second will be the complex conjugate of that for the first.

Having an asymptotic solution that is valid both near the singularity
and all the way to the interface, we may then impose the solvability
condition of vanishing mismatch function on this solution
\footnote{Actually, what is important is not that the solution remains
  a good approximation near the interface but only that it captures
  the transcendental term which in regular perturbation theory lies
  beyond all orders.}.

In the general case, we cannot simply factorize the basic partial
differential equation but achieve the reduction of order enabling
analytic solutions by Zauderer decomposition. Analytic continuation to
the complex plane will again prove useful. A convenient starting point
consists in representing the partial differential equations as a set
of first order equations. We define
\begin{align}
  \vecw &=\begin{pmatrix}
         T^l_\xi \\ \rule{0mm}{4mm} T^l_\eta
       \end{pmatrix} \>, \qquad 
       \vecw^s =\begin{pmatrix} T^s_\xi \\
        \rule{0mm}{4mm} T^s_\eta
       \end{pmatrix}\>, \qquad 
        \vecv =\begin{pmatrix}
         \psi_\xi \\
         \psi_\eta
        \end{pmatrix}\>,
\nonumber\\
A&=\begin{pmatrix}
   0 & 1 \\ -1 & 0
  \end{pmatrix}\>, \qquad  
B=\begin{pmatrix}
   -u_0 & -v_0 \\ 0 & 0
  \end{pmatrix}\>,  \qquad  
C=\begin{pmatrix}
   F_0 & 0 \\ 0 & 0
  \end{pmatrix}\>, 
\nonumber\\
u_0&=\psi_\eta+\xi P_f \>,\qquad
v_0=-\psi_\xi-P_f\lr{\eta-1}\>,\qquad \nonumber\\
F_0&=-\EXP{-\frac{P_f}{2}\lr{\eta-1}^2}\>.
\end{align}

The governing equations then become
\begin{subequations}
\begin{align}
 \vecw_\xi+A\vecw_\eta+B\vecw+C\vecv&=0 
\label{eq:w_eq}\\
 \vecw^s_\xi+A\vecw^s_\eta&=0 
\label{eq:ws_eq}\\
 \vecv_\xi+A\vecv_\eta&=0
\label{eq:v_eq}
\end{align}
\label{eq:first_order_system}
\end{subequations}
$A$ is a constant matrix, $B$ and $C$ are assumed to vary slowly as
functions of $\eta$ and $\xi$ in the vicinity of the Kruskal-Segur
point ($w=-\I$). This suggests a scale transformation
$\xi,\,\eta\to\varepsilon\xi,\,\varepsilon\eta$, emphasizing the
derivative terms in \eqref{eq:first_order_system}. As discussed in
\cite{fischaleck08}, $\varepsilon$ may be related to the stability
parameter after solution of the selection problem, giving
$\varepsilon\propto \sigma^{2/7}$.  We will expand equations in powers
of $\varepsilon$, drop terms of order $\varepsilon^2$ and higher and
set $\varepsilon=1$ afterwards to simplify the notation. A key of
Zauderer's approach is to rewrite the system of equations in terms of
eigenvectors of the matrix $A$ appearing in its principal part (here
given by expressions of the form $\vec f_\xi + A \vec f_\eta$) and to
obtain decoupled equations for their coefficients, using appropriate
projections onto the eigenvectors.  The eigenvectors of $A$ are
\begin{align}
\vecr_1 = \begin{pmatrix}
           -\I \\ 1
          \end{pmatrix}\>,\qquad
\vecr_2=\begin{pmatrix}
           \I \\ 1
          \end{pmatrix}\>, 
\end{align}
corresponding to the eigenvalues $\I$ and $-\I$, respectively. Since
$A$ is antihermitean and the eigenvalues different, these eigenvectors
are orthogonal (but the formalism does not rely on this). We expand
the field vectors in terms of $\vecr_1$ and $\vecr_2$
\begin{subequations}\label{eq:allexpans}
\begin{align}
 \vecw&=M\vec r_1+\varepsilon N\vec r_2\>,
\label{eq:expans_w}\\
 \vecw^s&=N^s\vec r_2\>,
\label{eq:expans_ws}\\
 \vecv&=\chi\vec r_1\>,\label{eq:expans_v}
\end{align}
\end{subequations}
where the choice of a prefactor $\varepsilon$ in front of the
coefficient function $N$ is dictated by our expectation of this term
being small in the liquid, because $\vecr_1$ is the eigenvector
leading to an equation of the form $M_\xi+\I M_\eta =0$ in the limit
$\varepsilon\to0$ that can be identified with the flowless case (see
Eq.~\eqref{eq:w_eq}, where the $B$ and $C$ terms become negligible
after the scale transformation in the limit $\varepsilon\to0$).  For
the case without flow, we have identified this form to correspond to
the equation generating the relevant component of our solution in the
vicinity of the singularity $w=-\I$. With flow, there will be
corrections of order $\varepsilon$ that we wish to calculate. That we
have completely dropped one of the eigenvectors in
Eqs.~\eqref{eq:expans_ws} and \eqref{eq:expans_v} is due to the fact
that Eqs.~\eqref{eq:ws_eq} and \eqref{eq:v_eq} have only principal
parts, so the coefficients of the dropped eigenvectors decouple
completely.  Equation~\eqref{eq:ws_eq} holds in the solid, so we
expect the $\vecr_2$ term to be dominant, Eq.~\eqref{eq:v_eq} refers
to the liquid domain, so the $\vecr_1$ term should be dominant.  Once
we have calculated the coefficient functions $M$, $N$, $N^s$, and
$\chi$, we may obtain the temperature and flow fields from
\begin{subequations}
\label{eq:old_by_new_fields}
\begin{align}
 T_\xi&=-\I\lr{M-\varepsilon N}\>, &
 T_\eta&=M+\varepsilon N\>, \\
 T_\xi^s&=\I N^s\>, &
 T_\eta^s&=N^s \>,\\
 \psi_\xi&=-\I\chi\>, & \psi_\eta &= \chi\>,
\end{align}
\end{subequations}
equations that also allow us to obtain boundary conditions for the
coefficient functions from Eqs.~\eqref{eq:interf_bcall}.

Plugging Eq.~\eqref{eq:allexpans} into
Eq.~\eqref{eq:first_order_system} and neglecting terms of order
$\varepsilon^2$, we find
\begin{subequations}
\label{eq:zauderer_vec}
\begin{align}
 M_\xi\vecr_1+\varepsilon N_\xi\vecr_2+\I M_\eta\vecr_1
-\I\varepsilon N_\eta\vecr_2& \nonumber\\
+\varepsilon BM\vecr_1
+\varepsilon C\chi\vecr_1&=0\label{eq:diffliq4}\\
 N^s_\xi-\I N^s_\eta&=0 \label{eq:diffsol1}\\
 \chi_\xi+\I \chi_\eta&=0\label{eq:psilaplace2}.
\end{align}
\end{subequations}

Next, we project these equations onto the eigenvectors to cast them in
the simplest possible scalar form. Projection operators on the two
eigenvectors are easily constructed by tensorial multiplication with
the dual vectors of the biorthogonal system constructed from $\vecr_1$ and $\vecr_2$.
 We have 
\begin{equation}\begin{aligned} 
P_1=\frac 12\begin{pmatrix}
           1 & -\I \\ \I & 1
          \end{pmatrix}\>,\quad
P_2&=\frac 12 \begin{pmatrix}
           1 & \I \\ -\I & 1
          \end{pmatrix}\>, 
\\
 P_i \vecr_k &= \delta_{i,k} \vecr_k\>,\quad i,k = 1,2\>.
\end{aligned}
\end{equation}
Abbreviating $a=\frac12\left(u_0+\I v_0\right)$, we may write the
projections of $B\vecr_i$ and $C\vecr_i$ as
\begin{equation}\begin{aligned} 
P_1 B \vecr_1 &= -a \vecr_1\>, &P_1 C \vecr_1 &= \frac{F_0}{2} \vecr_1 \>,\\
P_2 B \vecr_1 &= a \vecr_2\>, &P_2 C \vecr_1 &= -\frac{F_0}{2} \vecr_2\>.
\end{aligned}
\end{equation}
Applying the projection operators to \eqref{eq:diffliq4} and
setting $\varepsilon=1$, we obtain
\begin{subequations}
\label{eq:mn_equations}
\begin{align}
 M_\xi+\I M_\eta-aM+\frac{F_0}{2}\chi&=0\label{eq:diffliq51}\>,\\
 N_\xi-\I N_\eta+aM-\frac{F_0}{2}\chi&=0\label{eq:diffliq52}\>,
\end{align}
\end{subequations}
which together with \eqref{eq:diffsol1} and \eqref{eq:psilaplace2}
gives us four equations for the four quantities $M$, $N$, $N_s$, and
$\chi$.

Finally, the boundary conditions \eqref{eq:interf_bcall} have to be
transformed into boundary conditions for our new fields. Because
$\vecw$, $\vecw^s$, and $\vecv$ are defined in terms of derivatives of
the temperature field and the stream function, we have to take
derivatives in \eqref{eq:psilaplace2}, wherever non-differentiated
temperatures appear. It is here, where care has to be taken that the
boundary conditions hold along the interface and hence we do not
obtain a boundary condition for $T_\xi$ directly from \eqref{eq:cont1}
or \eqref{eq:Gibbs1} but one for $\D T/\D \xi=T_\xi+h'T_\eta$. Using
\eqref{eq:old_by_new_fields}, we find the interface conditions:
\begin{subequations}
\label{eq:zaud_interf_bc}
\begin{align}
 M&=\frac{\I}{2}\frac{\lre{\lr{1+\I\xi}h}'}{1+\I h'}\>,\label{eq:cond1}\\
 N-N^s&=-\frac{\I}{2}\frac{\lre{\lr{1-\I\xi}h}'}{1-\I h'}\>,\label{eq:cond2}\\
 N^s&=\frac{\I}{2}\frac{\sigma(\kappa a(\theta))'}{1-\I h'}\>,\label{eq:cond3}\\
 \chi&=-\frac{\I P_f\lre{\xi h}'}{1+\I h'}\>,\label{eq:masscon2}
\end{align}
\end{subequations}
where the prime always denotes a derivative with respect to $\xi$.
Combining the second and third equations, we see
that the equations for $M$ and $N$ decouple from that for $N^s$,
because we can give their boundary condition at the interface without
solving the equation for $N^s$ explicitly. (Of course, we have to make
sure that there \emph{is} a solution in the solid, so the behavior of $N^s$
near the singularity in question is important.) 

The boundary conditions at infinity follow from \eqref{eq:bc_inf2}
combined with \eqref{eq:field_linearization} and simply require all
fields to go to zero in the appropriate infinite limit:
\begin{subequations}
\label{eq:zaud_bc_inf}
\begin{align}
 M&\to 0  && (\eta \to \infty)\>, \label{eq:bnd_m_inf}\\
 N&\to 0  && (\eta \to \infty)\>,\label{eq:bnd_n_inf}\\
 N^s &\to 0  &&  (\eta<1,\> \abs{\xi} \to \infty)\>,\label{eq:bnd_ns_inf}\\
 \chi&\to 0  && (\eta \to \infty)\>.\label{eq:bnd_chi_inf}
\end{align}
\end{subequations}
The alert reader may be surprised that we have eight boundary
conditions [Eqs.~\eqref{eq:zaud_interf_bc} and \eqref{eq:zaud_bc_inf}]
for four first-order differential equations
[Eqs.~\eqref{eq:mn_equations}, \eqref{eq:diffsol1}, and
\eqref{eq:psilaplace2}]. The system looks heavily overdetermined.
Normally, this problem does not arise.  If the Zauderer method is applied to
a typical boundary value problem, for example, solving the Laplace
equation with Dirichlet boundary conditions on part of the boundary
and Neumann conditions on the remainder \footnote{In the form we
  employ, the method is not suited for Dirichlet boundary
  conditions, due to the transformation to a first-order system.},
then we will obtain boundary conditions for some combinations of
variables at the first boundary and for others at the second, with the
total number of conditions just corresponding to the total number of
equations.  However, our problem is not typical, as is well-known. The
interface position itself is an unknown of the problem, requiring the
imposition of an additional boundary condition at the interface. As a
consequence, we obtain a full set of boundary conditions already from
\eqref{eq:zaud_interf_bc}, but it is in terms of the unknown interface
position $h(\xi)$. The remaining boundary conditions
\eqref{eq:zaud_bc_inf} then are solvability conditions to be imposed
on that unknown function. It will turn out that three of these
conditions can be satisfied automatically by requiring $h(\xi)$ or the curvature to
approach zero sufficiently fast at infinity. The last one is a non-trivial
equation for $h(\xi)$ which replaces the integro-differential equation
derivable in problems with linear bulk equations. Applying the
Kruskal-Segur method to this interface equation, we may then derive
the selection equations.

\section{Solution of the decomposed equations}
\label{sec:sol_decompos}

All equations to be solved are now first order with linear derivative
terms. This suggests to try their analytic solution via the method of
characteristics, a step allowing to make progress
despite the nonlinearity of the basic equations.  

The principal parts of Eqs.~\eqref{eq:zauderer_vec} correspond to two sets
of characteristic coordinates. We start with \eqref{eq:psilaplace2}
and \eqref{eq:diffliq51}, first setting $\chi=\chi(s,\tau)$ with
$s=s(\xi,\eta)$ and $\tau=\tau(\xi,\eta)$. The linear combination of
derivatives should correspond to a derivative with respect to $s$
only, which yields the characteristic equations
\begin{align}
 \abl{\xi}{s}=1\>,\quad\abl{\eta}{s}=\I\>,
\quad
\abl{\chi}{s}=0\>.
\end{align}
Solving this system with the initial condition 
\begin{align}
\eta(s=0)=1\>, \quad \xi(s=0)=\tau
\end{align}
we obtain
\begin{subequations} 
\begin{align}
 s&=-\I\lr{\eta-1}\>,\\
 \tau&=\xi+\I\lr{\eta-1}\>,\\
\chi&=\chi(\tau) = -\frac{\I P_f\lre{\tau h(\tau)}'}{1+\I h'}\>,
\end{align}
\end{subequations}
i.e., $\chi$ is simply the analytic continuation into the upper $\eta$
half plane of the function represented by the boundary condition at
$\eta=1$ \eqref{eq:masscon2}. This was to be expected, since
Eq.~\eqref{eq:psilaplace2} contains only a principal part. Analyticity
requires $\chi$ to remain bounded for $\eta\to\infty$; in fact we have
the stronger condition \eqref{eq:bnd_chi_inf}. To make sure it is
satisfied, we may impose  the perturbation $h(\tau)$ to decay fast
enough for $\tau\to \I\infty$ so that $\lre{\tau h(\tau)}'\to 0$.

The case of the function $M$ is more interesting. We have the same
characteristic coordinates $s$ and $\tau$, and the equation for $M$
takes the form
\begin{align}
 M_s-\frac{P_f}{2}\lr{2s+\tau}M+\frac{\I P_f\lre{\tau 
h(\tau)}'}{2\lr{1+\I h'}}\EXP{\frac{P_f}{2}s^2}=0
\label{eq:charact_M}\>.
\end{align}
Solving this with initial condition \eqref{eq:cond1}, we find
\begin{align}
  M(s,\tau)=\frac{\I}{2}&\EXP{\frac{P_f}{2}\lr{s^2+s\tau}}
\left[\rule{0pt}{5mm}\right.\frac{\lre{\lr{1+
        \I\tau}h}'}{\lr{1+\I h'}}
  \nonumber\\
  &-\frac{2\lre{\tau h}'}{\tau\lr{1+\I
      h'}}\lr{1-\EXP{-\frac{P_f}{2}s\tau}}
  \left.\rule{0pt}{5mm}\right]
\label{eq:solution_m}
\end{align}
and boundary condition \eqref{eq:bnd_m_inf} is satisfied, if again we
assume $h(\tau)\to 0$ for $\tau\to\I\infty$.

The characteristic coordinates for the other two equations are 
\begin{align}
 \bar s&=\I\lr{\eta-1}\>,\quad
 \bar\tau=\xi-\I\lr{\eta-1}
\end{align}
and we obtain 
\begin{align}
\abl{N^s}{\bar s} = 0 \>,
\end{align}
giving the obvious solution
\begin{align}
N^s&=\frac{\I}{2}\frac{\sigma\lre{\kappa(\bar \tau) a(\theta)}'}{1-\I h'(\bar \tau)}\>,
\label{eq:solution_ns}
\end{align}
and boundary condition \eqref{eq:bnd_ns_inf} is satisfied, if we
require the (derivative of the) curvature term $\kappa a$ to vanish for
$\bar\tau\to-\I\infty$ \footnote{Note that the curvature vanishes for
  $h\to\infty$ and becomes equal to the curvature of the Ivantsov
  parabola for $h\to 0$, hence vanishes for
  \protect{$\vert\xi\vert\to\infty$} in that case.}.

Finally, the equation for $N$ becomes 
\begin{align}
N_{\bar s}=-\frac{P_f}{2}\bar\tau M(-\bar s,\bar\tau+2\bar s)+\frac{\I P_f\lre{(\bar\tau
+2\bar s)h(\bar\tau+2\bar s)}'}{2\lr{1+\I h'(\bar\tau+2\bar s)}}\EXP{\frac{P_f}{2}\bar s^2}
\label{eq:equation_n}
\end{align}
with the boundary condition at $\bar s=0$, following from
\eqref{eq:cond2} and \eqref{eq:cond3} with $\xi=\bar\tau$:
\begin{align}
 N(\bar s=0)=\frac{\I}{2}\frac{\sigma(\kappa a)'(\bar\tau)}{1-\I h'(\bar\tau)}
-\frac{\I}{2}\frac{\lre{\lr{1-\I\bar\tau}h(\bar\tau)}'}{1-\I h'(\bar\tau)}.
\end{align}
Equation \eqref{eq:equation_n} can be solved by direct quadrature, with the result: 
 \begin{align}
   N(\bar s,\bar\tau)=&-\frac{P_f}{2}\bar\tau\int\limits_0^{\bar s}
M\lr{-\bar s',\bar\tau+2\bar s'}\D\bar s'
\nonumber\\
   &+\frac{\I P_f}{2}\int\limits_0^{\bar s}\frac{\lre{\lr{\bar\tau+2\bar s'}
h(\bar\tau+2\bar s')}'}{1+\I h'(\bar\tau+2\bar s')}\EXP{\frac{P_f}{2}\bar s'^2}\D\bar s'
\nonumber\\
   &+\frac{\I}{2}\lre{\frac{\sigma(\kappa a)'(\bar\tau)}{1-\I
         h'(\bar\tau)}-\frac{\lre{\lr{1-\I\bar\tau}h(\bar\tau)}'}{1-\I
         h'(\bar\tau)}}
\label{eq:Nsol}
\end{align}
and a sufficient condition for boundary condition \eqref{eq:bnd_n_inf}
to be satisfied is
\begin{align}
 \lim\limits_{\bar s\to\I\infty}N(\bar s,\bar\tau)=0.
\label{eq:limN_sbartoinf}
\end{align}
Evaluation of this requirement will produce the central equation, to
which the Kruskal-Segur method can be applied. Note that
Eq.~\eqref{eq:limN_sbartoinf} is an equation for the interface
position $h(\bar \tau)$ that has to be  satisfied identically in the
single complex variable $\bar\tau$. The next task is to cast this
equation into a useful form. Since this is purely technical, we
relegate the calculation to the appendix. The resulting interface
equation is
\\[-12mm]
\begin{widetext}
\begin{equation}
 \begin{aligned}
   \sigma\kappa(\xi) a(\xi)=&\,\lr{1-\I\xi}h(\xi)+\frac{P_f}{4}\EXP{\frac{P_f}{8}\xi^2}\int\limits^\xi\EXP{-\frac{P_f}{8}\xi'^2}\left[\rule{0pt}{.9cm}\frac{1-\I
       h'(\xi')}{1+\I h'(\xi')}\xi'\lr{1-\I\xi'}h(\xi')\right.
   \\
   &\left.+\frac{P_f}{2}\lr{1-\I
       h'(\xi')}\int\limits_{\xi'}^{\I\infty}\xi''h(\xi'')\frac{\xi''-\xi'}{1+\I
       h'(\xi'')}\EXP{\frac{P_f}{8}\lr{\xi'-\xi''}^2}\D\xi''\right]\D\xi'
   \\
   &-\frac{P_f}{2}\EXP{\frac{P_f}{8}\xi^2}\int\limits^\xi\EXP{-\frac{P_f}{8}\xi'^2}\left[\lr{1-\I
       h'(\xi')}\int\limits_{\xi'}^{\I\infty}\xi''h(\xi'')\frac{\I
       h''(\xi'')}{\lr{1+\I
         h'(\xi'')}^2}\EXP{\frac{P_f}{8}\lr{\xi'-\xi''}^2}\D\xi''\right.
   \\
   &\left.+\xi'\int\limits^{\xi'}h''(\xi'')\lr{\frac{\I\lr{1+\I\xi''}h(\xi'')}{\lr{1+\I
           h'(\xi'')}^2}+\int\limits_{\xi''}^{\I\infty}M\lr{\frac{1}{2}(\xi''-\xi'''),\xi'''}\D\xi'''}\D\xi''\right]\D\xi'\>,
 \end{aligned}
\label{eq:F3}
\end{equation}
\end{widetext}
where we have renamed $\bar\tau$ into $\xi$ for convenience and
replaced the argument $\theta$ of the anisotropy function also by
$\xi$ (the dependence $a(\xi)$ is given in the appendix).  Note that
we can immediately read off the limit of vanishing flow ($P_f\to 0$)
and verify that it agrees with the local equation
\eqref{eq:bigeq_simple}. [This may of course already be seen from
Eqs.~\eqref{eq:Nsol} and \eqref{eq:limN_sbartoinf}.] The full equation
is nonlocal but it is tractable via asymptotic methods. 


In principle, Zauderer's scheme may be used to solve the system of
equations \eqref{eq:first_order_system} perturbatively. To carry out
this (complicated) calculation, one would have to keep the dropped
terms of order $\varepsilon^2$ and to add terms containing the missing
eigenvectors (and a factor $\varepsilon$) to Eqs.~\eqref{eq:expans_ws}
and \eqref{eq:expans_v}.  Inspection immediately reveals that this
expansion in powers of $\varepsilon$ does not correspond to a
perturbation series about the flow-Ivantsov solution: setting
$\varepsilon=0$ does not give us the full flow-Ivantsov solution but
only the solutions of the first-order equations obtained from the
projection onto eigenvectors of $A$; e.g., in the case of the Laplace
equation for $\chi$ we would just obtain a solution to $\chi_\xi+\I
\chi_\eta=0$. However, these pieces of the full solution are the ones
that diverge in the limit $\sigma\to 0$ near the singularity of
interest, whereas the other terms remain finite. Hence, the
lowest-order Zauderer solution corresponds to the exact solution of
the problem near the singularity. So the perturbative scheme arising
from Zauderer decomposition corresponds to an expansion about the
analytic continuation of the flow-Ivantsov solution in the vicinity of
the singularity. This may be seen as the deeper reason why a condition
for the transcendental term which is beyond all orders in regular
perturbation theory appears already at first order in our approach --
near the singularity this term is not small, so it has to be present
in a Zauderer type perturbation theory.

We have carried this perturbative approach beyond first order for the
simpler problem without flow. If we expand $h$ and $\kappa$ in powers
of $\varepsilon$ as well, i.e., $h=h_0+\varepsilon h_1+\ldots$ and
$\kappa = \kappa_0+\varepsilon \kappa_1+\ldots$, a solvability
condition similar to \eqref{eq:bigeq_simple} appears to turn up at the
next order involving $h_1$ and $\kappa_1$ -- but it is automatically
satisfied. Hence, it seems that the lowest-order solvability condition
does indeed capture the mismatch function needed to obtain
the selection criterion. While we knew this to be true from comparison
with known results in the case of \eqref{eq:bigeq_simple}, these
arguments suggest it to hold in general, i.e., also for
Eq.~\eqref{eq:F3}.

In order to obtain the mismatch function (or the contribution to it by
the singularity considered) at the interface, we have to solve
Eq.~\eqref{eq:F3}. Far from the singularity, this can be done by
linearization in terms of the interface position $h$ and its
derivatives, which may all be considered small. The appropriate tool
is WKB analysis.  Due to the linearity of the problem, this will
provide the solution up to a constant factor only. Using
asymptotic analysis, we can then solve Eq.~\eqref{eq:F3} near the
singularity, taking all important nonlinearities into account.
Asymptotic matching of the two solutions provides the prefactor of the
near-interface solution. The mismatch function calculated from it must
vanish at the tip of the needle crystal -- this is the solvability
condition.

\section{WKB analysis far from the singularity}
\label{sec:wkb}

Linearizing ~\eqref{eq:F3}, we obtain the inhomogeneous linear equation\\
\begin{align}
&\sigma\lr{\frac{1}{\lr{1+\xi^2}^{\frac 32}}-\frac{h''(\xi)}{\sqrt{1+\xi^2}}
-\frac{\xi h'(\xi)}{\lr{1+\xi^2}^{\frac 32}}}
=\lr{1-\I\xi}h(\xi)
\nonumber\\
&\quad+\frac{P_f}{4}\EXP{\frac{P_f}{8}\xi^2}\int\limits^\xi
\EXP{-\frac{P_f}{8}\xi'^2}\xi'\lr{1-\I\xi'}h(\xi')\D\xi'\>.\label{eq:WKB1} 
\end{align}
The solution of this consists of a particular solution to the
inhomogeneous equation (which will be captured by regular perturbation
theory) plus the general solution of the homogeneous equation (with
integration constants to be determined from boundary conditions on
$h$). The latter consists of an exponentially small and an
exponentially large term. The large term is suppressed already within
regular perturbation theory, but the small one will not appear therein
at any finite order. It becomes important, when, due to symmetries of
the problem, all terms of regular perturbation theory vanish. In the
needle-crystal problem, this is the case at the tip of the crystal. So
the transcendentally small term that must be suppressed can be
identified with the decaying solution of the homogeneous linear
equation corresponding to \eqref{eq:WKB1}.  Alternatively, we could
argue that the general solution to the inhomogeneous equation may be
obtained, within WKB theory, via the method of variation of constants
\cite{bender78}. Again, the exponentially large term must be
eliminated by an appropriate choice of an integration parameter. The
exponentially small one has the same form as the decaying solution of
the homogeneous equation, except that there is now a slowly varying
prefactor. Since the mismatch function is to be evaluated at the tip
position in the end, it has the same form as this solution.

The only tricky part of the calculation of the WKB solution is the
evaluation of the integral in \eqref{eq:WKB1}, which can be done via
integration by parts. We obtain
\begin{align}
h(\xi)=B_1\EXP{\frac{P_f}{16}}\lr{1+\I\xi}^{-\frac 38}\lr{1-\I\xi}^{-\frac 58}
\EXP{\frac{S_0(\xi)}{\sqrt{\sigma}}+\frac{P_f}{16}\xi^2}
\label{eq:WKBsolution}
\end{align}
with an unknown constant $B_1$ and
\begin{equation}
 S_0(\xi)=\I\int\limits_{-\I}^\xi \lr{1+\I\xi'}^{\frac 14}\lr{1-\I\xi'}^{\frac 34}\D\xi'.
\end{equation}

\section{Solution near the singularity}
\label{sec:sol_near_sing}

The most appropriate form of Eq.~\eqref{eq:F3} for a local analysis near
the singularity seems to be Eq.~\eqref{eq:F2}. With $M$ given explicitly, it reads
\begin{align}
  F(\xi)=&\frac{P_f}{4}\int\limits^\xi\int\limits_{\xi'}^{\I\infty}
  \frac{\bar z(\xi')}{z(\xi'')}\left[\xi'
    \EXP{\frac{P_f}{8}\lr{\xi'^2-\xi''^2}}
    \left(\vphantom{\EXP{\frac{P_f}{4}}}\lre{\lr{1+\I\xi''}h(\xi'')}'\right.\right.
  \nonumber\\
  &-\frac{2}{\xi''}\lre{\xi''h(\xi'')}'
  \left.\left.\cdot\lr{1-\EXP{\frac{P_f}{4}\xi''\lr{\xi''-\xi'}}}\right)\right.
  \nonumber\\
  & \left.
    -2\lre{\xi''h(\xi'')}'\EXP{\frac{P_f}{8}\lr{\xi'-\xi''}^2}\right]
\D\xi''\D\xi'.\label{eq:F4}
\end{align}
Introducing the stretching transformation\footnote{It should be kept
  in mind that the variable $t$ introduced here has nothing to do with
  a time.  Nevertheless, we denote derivatives
  with respect to $t$ by overdots.}
\begin{equation}
  \xi=-\I\lr{1-\sigma^\alpha t}
\end{equation}
with $\alpha=\frac27$ (obtained from a dominant balance consideration), we set
\begin{equation}
h(\xi)=\sigma^\alpha\phi(t)
\end{equation}
from which we get
\begin{subequations}
 \begin{align}
h'&=-\I\dot\phi
\\
h''&=-\sigma^{-\alpha}\ddot\phi
\\
\lr{1-\I\xi}h&=\sigma^{2\alpha}\phi t
 \end{align}
\end{subequations}
and find, to leading order in $\sigma$
\begin{align}
F=\sigma^{2\alpha}&\left[\rule{0pt}{8mm}\right.\frac{1}{\sqrt{2t+2\phi}}
\left(\rule{0pt}{7.5mm}\right.\frac{\ddot\phi}{\lr{1-\dot\phi^2}^{\frac 32}}
\nonumber\\
&\quad+\frac{1+\dot\phi}{\lr{2t+2\phi}\sqrt{1-\dot\phi^2}}\left.\rule{0pt}{7.5mm}\right)-\phi t
\left.\rule{0pt}{8mm}\right].
\label{eq:scaling_of_F}
\end{align}
Rewriting the right-hand side of Eq.~\eqref{eq:F4} in terms of the
new variable is a bit more involved and leads to 
\begin{align}
F&=\frac{P_f}{4}\sigma^{3\alpha}
\int\limits^t\int\limits_\infty^{t'}\frac{1-\dot\phi(t')}{1+\dot\phi(t'')}
\left[\rule{0pt}{5mm}\right.\EXP{\frac{P_f}{4}\sigma^\alpha(t'-t'')}\lr{t''\dot\phi-\phi}
\nonumber\\
& +2\lr{t'-t''}\dot\phi\left.\rule{0pt}{5mm}\right]\D t''\D t'\>.\label{eq:F5}
\end{align}
An important result is that the leading order of $\sigma$ is
$\sigma^{3\alpha}$ which in the limit $\sigma\to 0$ vanishes faster
than the leading order of Eq.~\eqref{eq:scaling_of_F}. This means that
for finite $P_f$ the selected stability parameter will be the same as
in the flowless case with otherwise equal parameters. Hence, for the
same $P_c$, the same tip radius (and velocity) will be selected with
and without flow. For given undercooling, the selected velocity and
tip radius will be different from the corresponding quantity of the
flowless case only due to the different relationship
\eqref{eq:selection1} between undercooling and growth Péclet number,
which contains a dependency on $P_f$ (i.e., for the same $\Delta$,
$P_c$ is different in the two cases). In fact, this result has been
used as an \emph{assumption} in the past to compute selected growth
velocities in convective situations \cite{ananth90}. Here, it has been
proved for the case of potential flow, but our experience with other
flow patterns suggests this to be a general feature of convection. To
our knowledge, no general proof has been given so far.

To obtain a nontrivial dependency of the stability parameter on the
flow Péclet number, we have to assume large flow velocities,
e.g.~$P_f= \mathcal{O}(\sigma^{-\alpha})=\mathcal{O}(\sigma^{-2/7})$. Hence we set
\begin{align}
P_1=\frac{P_f}{4}\sigma^{\alpha}.
\end{align}

Since we expect that there is no solution in the isotropic case, we
take surface tension anisotropy into account right away. Performing
the stretching transformation for $a(\theta)$, we find
\begin{align}
  a(\theta) = 1-
  \frac{2\beta\sigma^{-2\alpha}\lr{1-\dot\phi}^2}{\lr{t+\phi}^2\lr{1+\dot\phi}^2}\>.
\label{eq:aniso_stretch}
\end{align}
The anisotropy parameter $\beta$ usually is numerically small. Setting
\begin{align}
\beta = \sigma^{2\alpha} b = \sigma^{4/7} b\>,
\end{align}
we may cast our interface equation into the form
\begin{align}
\phi t&+P_1\int\limits^t\int\limits_\infty^{t'}\frac{1-\dot\phi(t')}{1+\dot\phi(t'')}
\left[\rule{0pt}{5mm}\right.\EXP{P_1(t'-t'')}\lr{t''\dot\phi-\phi}
\nonumber\\
&\qquad\qquad+2\lr{t'-t''}\dot\phi\left.\rule{0pt}{5mm}\right]\D t''\D t'=
\nonumber\\
&\frac{1}{\sqrt{2t+2\phi}}\lre{\frac{\ddot\phi}{\lr{1-\dot\phi^2}^{\frac 32}}
+\frac{1+\dot\phi}{\lr{2t+2\phi}\sqrt{1-\dot\phi^2}}}
\nonumber\\
&\qquad\quad\times\lre{1-\frac{2b\lr{1-\dot\phi}^2}{\lr{t+\phi}^2\lr{1+\dot\phi}^2}}\>.
\label{eq:F6}
\end{align}
Given the boundary condition that the imaginary part of $\dot \phi$
vanishes for $t\to\infty$ (which is the condition that the tip slope
of the needle crystal is equal to zero) and a prescribed value of
$P_1$, this constitutes a nonlinear eigenvalue problem for $b$. We
have solved this numerically in the complex plane, using a scheme
similar to the one given by Tanveer \cite{tanveer89}; we employ a
relaxation method along two straight intersecting lines in the complex
plane, one of them parallel to the imaginary, the other lying on the
real $t$ axis. Details of the numerical approach, which is a root
finding problem involving the integration of several ordinary
differential equations and exhibits a certain level of complexity, will
be given elsewhere.

We do not find any solutions with $b=0$, suggesting that there does
not exist, as anticipated, any steady-state needle crystal close to a
flow-Ivantsov parabola in the case of isotropic surface tension. 

For anisotropic surface tension, we have the usual relationship
between the selected stability parameter and the anisotropy parameter
\begin{equation}
\sigma=\lr{\frac{\beta}{b}}^{\frac 74}\>.
\end{equation}
If the general solution behavior is similar to that of the flowless
case, the solution corresponding to the lowest eigenvalue $b$ should
be the only one that is linearly stable. We assume this to be true,
but have not yet been able to check it. 

The relationship between the physical flow Péclet number and our $P_1$ is
\begin{equation}
P_f=4P_1\sqrt{\frac{b}{\beta}} \>.
\end{equation}
Once we have $P_f$ and $\sigma$, we may determine $P_c$ (numerically
\footnote{Since our calculation is valid for $P_c\ll 1$, we may also,
  for finite $P_f$, use the analytic approximation obtained by setting
  $P_c=0$ on the right-hand side of Eq.~\eqref{eq:selection1}, without
  changing the order in $P_c$ up to which the calculation is correct;
  in the case of $P_f=0$, we first have to evaluate the integral on
  the right-hand side, but still may set $P_c=0$ in the exponential
  prefactor. The numerical evaluation of $P_c$ interpolates smoothly
  between these two limits, corresponding to $P_c\propto \Delta$ and
  $P_c\propto \Delta^2$, respectively.})  from
Eq.~\eqref{eq:selection1} and using the definitions
\eqref{eq:def_growpeclet} of $P_c$ and \eqref{eq:def_sigma} of
$\sigma$ we can evaluate both the selected tip radius $\rho$ and tip
velocity $V$.

Note that while our approximations hold in the limit $P_c\ll P_f$,
which implies in particular an approximation for $T^l$ in
Eq.~\eqref{eq:potTIv2} that does not approach the limit $P_f\to 0$
uniformly in $\eta$, the eigenvalue $b$ obtained numerically will
still be correct in that limit, due to the structure of
Eq.~\eqref{eq:F3} which reduces to the selection criterion without
flow. Indeed, we have verified that we obtain the same value of $b$ as
Tanveer \cite{tanveer89} in the case without flow.

Although our model is definitely a toy model -- experimental flow
patterns and velocities will not be well described by a potential flow
\footnote{A potential flow would be expected around solid helium
  growing into its superfluid. For such a system, the Gibbs-Thomson
  condition will not describe the interface temperature correctly
  anymore due to the appearance of a Kapitza resistance. Moreover, the
  only experiments on dendritic growth with solid helium we are aware
  of \protect{\cite{franck86,rolley94}} ($^4$He, $^3$He) were done at
  temperatures well above the transition to superfluidity.} -- we
carry the calculation to its end using parameters determined for an
experimental substance, pivalic acid. Since it is not to be expected
that this will give more than qualitative trends, the purpose of this
exercise is mostly to demonstrate that the (relatively elaborate)
formalism produces numbers finally and that these numbers do not have
unreasonable orders of magnitude.

Caveats to be kept in mind are: \\ 
-- We use the symmetric model, whereas the one-sided model would be
more appropriate for experiments with solute diffusion.  However, this
is known to just make a difference of a factor of two in the selected
velocity \cite{misbah87} in the diffusion-limited case.  We expect a
similar closeness of results of the two models in the presence of
convection. \\
-- Our model is only two-dimensional, which certainly impedes its
quantitative applicability to experiments. On the other hand,
typically the predictions of microscopic solvability theory do not
differ much for two-dimensional and (axisymmetric) three-dimensional
systems \cite{muschol92}.\\
-- More importantly, pivalic acid has kinetic anisotropy, so it is not
to be expected anyway that a model imposing local equilibrium at the
interface will yield a good description.  We chose the experiments
from Ref.~\onlinecite{bouissou89} for comparison, because they have
flow velocities that are in the range of numerical accessibility for
our code, whereas in experiments with succinonitrile \cite{lee93} (a
system expected to be better suited for comparison on physical
grounds), the imposed flow velocities were very large, leading to
convergence problems in our eigenvalue computation.\\  
-- Potential flow
and hence our relationship between $P_c$ and $P_f$ is not realized in
the experiments.

\begin{figure}[ht!]
\includegraphics[width=0.47\textwidth]{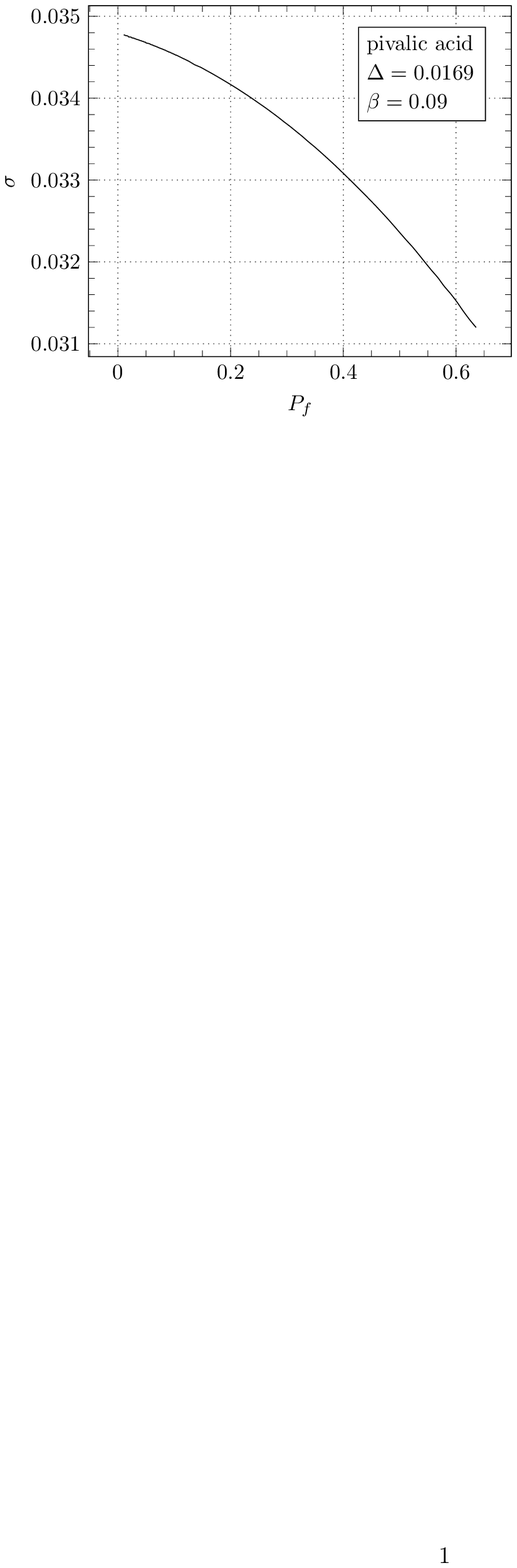}
\caption{The stability parameter $\sigma$ as a function of the flow
 P\'eclet number $P_f$. Material parameters used correspond to pivalic acid with
   $\Delta=0.0169$ and $\beta=0.08$ \protect{\cite{dougherty91}}. }
\label{fig:sig_of_Pf}
\end{figure}

\begin{figure}[h!]
\includegraphics[width=0.47\textwidth]{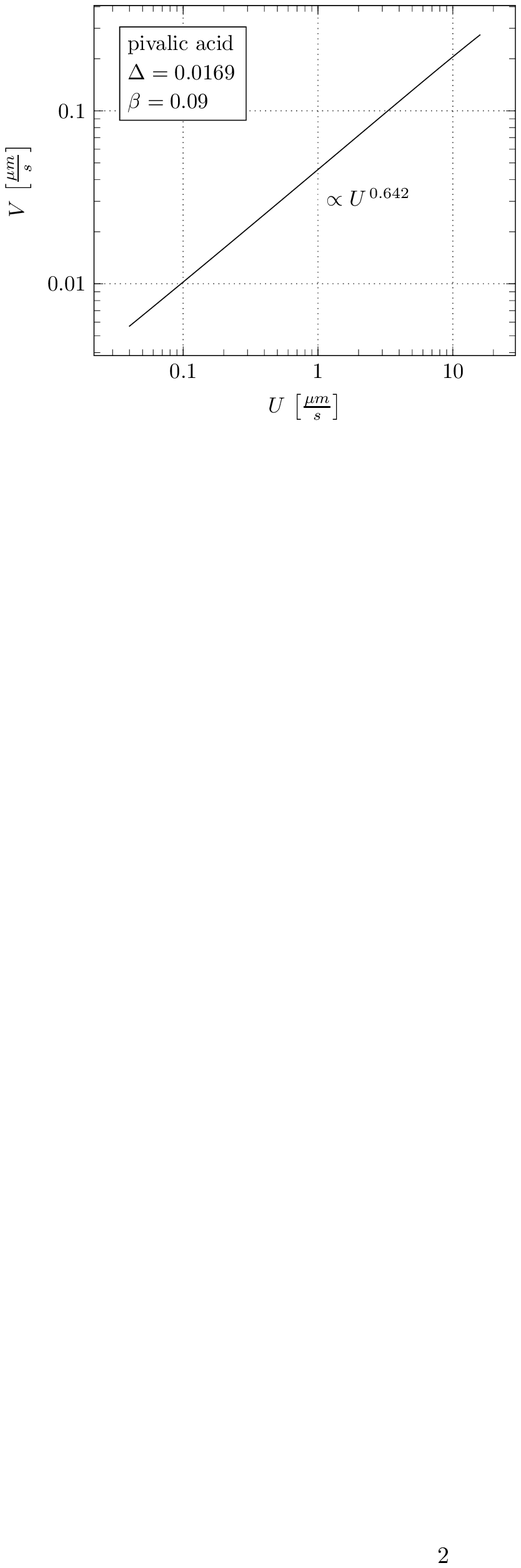}
\caption{The crystal growth velocity $V$ as a function of the flow
  velocity $U$.}
\label{fig:V_of_U}
\end{figure}

\begin{figure}[h!]
\includegraphics[width=0.47\textwidth]{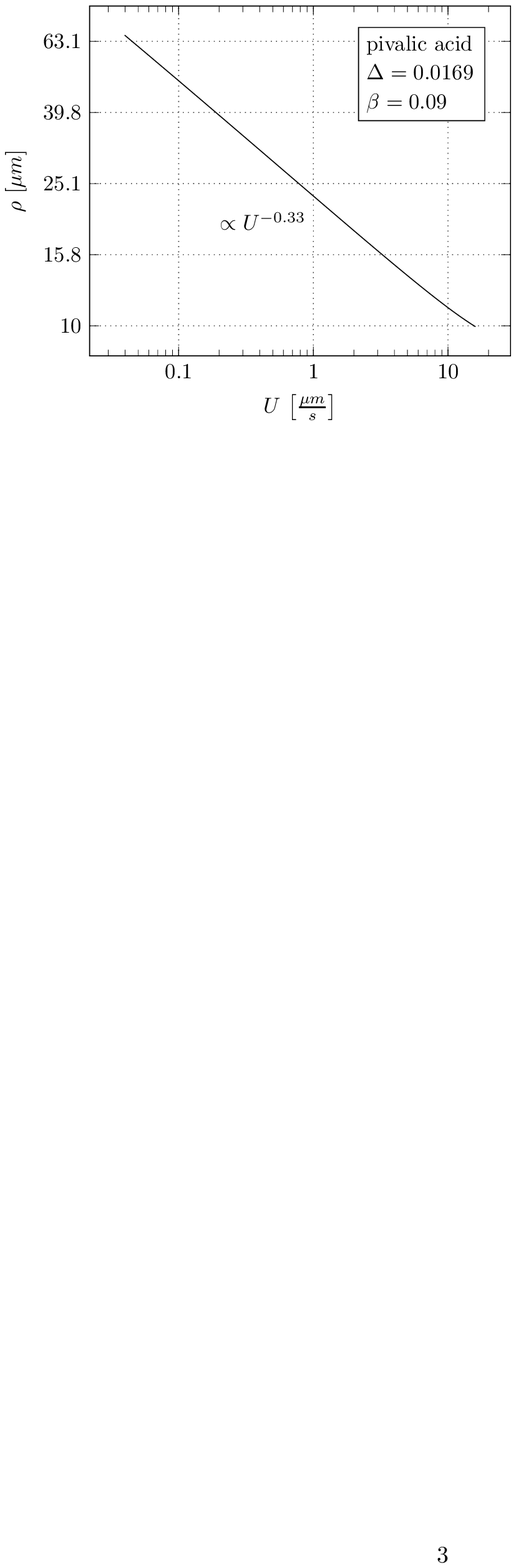}
\caption{The tip radius of the needle crystal $\rho$ as a function of
  the flow velocity $U$.}
\label{fig:rho_of_U}
\end{figure}

\begin{figure}[h!]
\includegraphics[width=0.47\textwidth]{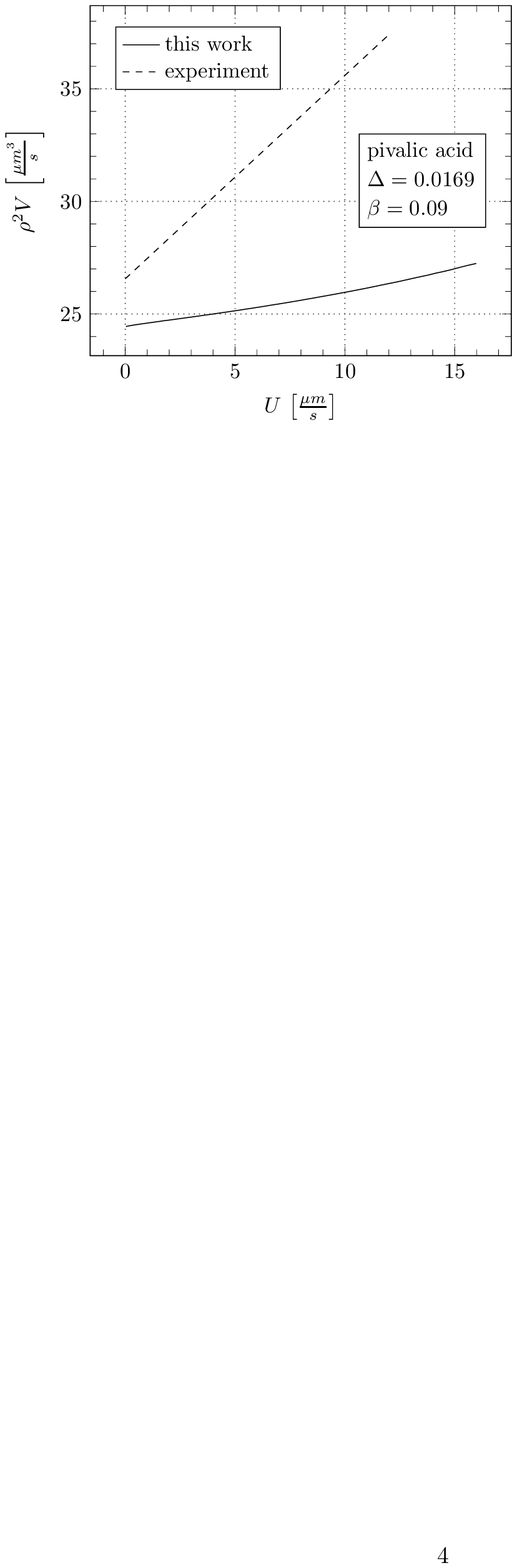}
\caption{The product $\rho^2 V$ as a function of the flow velocity
  $U$, compared with an experiment \protect{\cite{bouissou89}}.}
\label{fig:rho2V_of_U}
\end{figure}

Material parameters were taken from Refs.~\onlinecite{bouissou89},
\onlinecite{RG91}, and \onlinecite{dougherty91} and an undercooling of
about 0.2 K (equivalent to $\Delta=0.0169$) was assumed,
corresponding to a situation considered in the experiments.  Results
are shown in Figs.~\ref{fig:sig_of_Pf} to \ref{fig:rho2V_of_U}.

Figure~\ref{fig:sig_of_Pf} gives the selected value of $\sigma$ as a
function of the flow Péclet number for fixed undercooling $\Delta$
and an anisotropy parameter that corresponds to a measured value 
\cite{dougherty91}.

In Figs.~\ref{fig:V_of_U} and \ref{fig:rho_of_U} we give the selected
growth velocity $V$ and tip radius $\rho$ \footnote{This is not the
  radius of curvature at the tip of the true crystal but the one
  corresponding to a flow-Ivantsov solution traveling at the same
  velocity, i.e., the radius should be obtained by fitting the tail of
  an experimental needle crystal -- after removal of noise-induced
  side branches -- to a parabola. Since correction for side branches
  is tricky, one may instead fit to an appropriate piece of the needle
  crystal ahead of the side-branching region but not too close to the
  tip.} in dimensional form. We refrain from comparing these
numerical results  with a concrete experiment,
because there are too many uncertainties regarding the applicability
of the toy model to real life.

All that we wish to point out here is that there are power law scaling
relations between the growth velocity and the velocity of the imposed
flow as well as between the tip radius and the flow velocity, valid in
a range of undercoolings. This feature will probably not disappear in
a more quantitative calculation. In fact, we have checked for an
extended range of anisotropies, thus varying $\sigma$ between small and
very large values, that the scaling exponents change only slightly.

Finally, we do  compare the values of $\rho^2 V$ obtained from this
calculation with experimental values in a flow situation
\cite{bouissou89}, because $\rho^2 V$ is expected to be a slowly
varying quantity and therefore what matters mostly is the overall
order of magnitude. As Fig.~\ref{fig:rho2V_of_U} demonstrates, this
quantity compares reasonably with experiment. In fact, considering
that the experimentalists describe their flow pattern as approximate
Oseen flow, the agreement is not too bad. This should of
course not be taken too seriously either. A real comparison will have
to await a calculation with a more realistic flow (and, for pivalic
acid, a different interface boundary condition).

To conclude this section, it may be noted that a local asymptotic analysis of
Eq.~\eqref{eq:F6} for $t\to\infty$ yields the same transcendental behavior
as Eq.~\eqref{eq:WKBsolution} and provides the prefactor $B_1$ in terms of
the solution of the nonlinear equation \eqref{eq:F6} as a function of
$P_1$ and $b$. Since the boundary condition on the tip slope was
however already incorporated into the numerical scheme for the
solution of Eq.~\eqref{eq:F6}, this calculation does not provide anything
new.

\section{Conclusions}
\label{sec:conclusions}

After introducing the combination of Zauderer decomposition with the
Kruskal-Segur approach recently \cite{fischaleck08,fischaleck08b}, we have now
presented the method in more detail. The analytic part of the
calculation has been exemplified with a fully nonlinear problem.
Approximations that were introduced in \cite{fischaleck08} for
didactic reasons have been removed, rendering the full power of
the method visible.

We believe our approach to be the only one presented so far that has
the potential of a rigorous solution of pattern selection problems
with nonlinear bulk equations. Essentially, our belief that the method
is rigorous rests on two facts: first, the Zauderer decomposition
scheme produces a solution that becomes exact near the appropriate
complex-plane singularity; second, the WKB solution derived from the
interface equation within the scheme generates the same transcendental
terms that a WKB solution derived from the full system of partial
differential equations would. The second statement has been shown to
be true for the flowless case \cite{fischaleck08b} and we have given
arguments here, why it should carry over to the nonlinear case as well.

The elegance and power of the method show up in its rendering the
purely diffusion-limited case almost trivial
\cite{fischaleck08,fischaleck08b}. When applied to a problem with
nonlinear bulk equations, calculations certainly become involved. But
the problem remains solvable in a controlled manner, not provided by
other methods.  That in the final step the numerical determination of
an eigenvalue becomes necessary should not prevent us from considering
the approach basically analytical. A similar final step is necessary
in almost all related problems with simpler bulk equations, even though the
nonlinear equation to be solved numerically is less difficult in these cases.

We are convinced that our method will render a number of selection
problems accessible to solvability theory for which controlled
approximations could not be developed in the past, thus opening a new
line of research. These problems would include nonlinear diffusion
\cite{kurtze87}, density-driven convection \cite{sun09} (for which we
have given a preliminary treatment before \cite{fischaleck02}), Oseen
flow \cite{emsellem95,bouissou89}, the effect of the Kapitza
resistance on dendritic growth of helium \cite{franck86,rolley94}, but
also pattern selection problems outside of crystal growth such as, for
example, the motion of the two-phase front between superconducting and
normal conducting parts of a material \cite{chapman95}.

{\bf Acknowledgments} Financial support of this work by the German
 Research Foundation (DFG) under grant no.~KA 672/10-1 is gratefully
 acknowledged.
\begin{appendix}
\section{Conformal parabolic coordinates}
\label{sec:conformal_parabolic}

The unnormalized coordinate basis is given by
\newcommand{\EE}{\boldsymbol{\mathcal{E}}} 
\begin{align}
\EE_\xi\equiv\pabl{\vec x}{\xi} = \begin{pmatrix} \eta \\ -\xi
\end{pmatrix}\>,
\qquad \EE_\eta\equiv\pabl{\vec x}{\eta} = \begin{pmatrix} \xi \\ \eta
\end{pmatrix}\>,
\end{align}
which yields $g_\xi^2 \equiv \abs{\EE_\xi}^2=\xi^2+\eta^2 = \abs{\EE_\eta}^2 \equiv g_\eta^2$.

\begin{figure}[h!]
\includegraphics[width=0.48\textwidth]{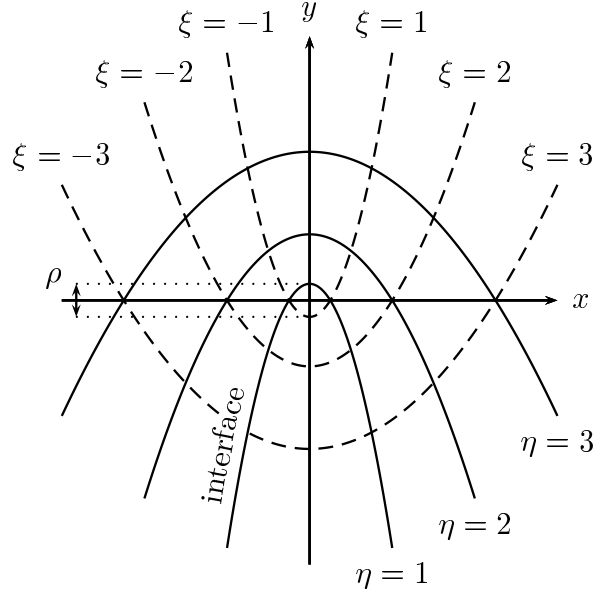}
\caption{Coordinate lines for conformal parabolic coordinates.
       Note that these coordinates will cover the $xy$ plane
      twice, if negative values for $\eta$ are admitted.}
\label{fig:coordinates}
\end{figure}

For the nabla operator, we get
\begin{equation}
 \nabla=\frac{1}{\sqrt{\xi^2+\eta^2}}\left[\vec e_\xi\pabl{}{\xi}+\vec e_\eta\pabl{}{\eta}\right]
\>,
\end{equation}
whereas the Laplacian reads
\begin{equation}
 \laplace=\frac{1}{\xi^2+\eta^2}\left[\pabl[2]{}{\xi}+\pabl[2]{}{\eta}\right].
\end{equation}
After normalization, the basis vectors are
\begin{equation}
\begin{aligned}
\vec e_\xi &= \frac{1}{\sqrt{\xi^2+\eta^2}}\left( \eta\vec e_x -\xi\vec e_y\right)
\>,\\
\vec e_\eta &= \frac{1}{\sqrt{\xi^2+\eta^2}}\left( \xi\vec e_x +\eta\vec e_y\right)
\>,
\end{aligned}
\end{equation}
which can be inverted to express the Cartesian basis by the orthonormal parabolic one
\begin{equation}
\begin{aligned}
 \vec e_x&=\frac{1}{\sqrt{\xi^2+\eta^2}}\left(\eta\vec e_\xi +\xi\vec e_\eta\right)\>,
\\
 \vec e_y&=\frac{1}{\sqrt{\xi^2+\eta^2}}\left(\eta\vec e_\eta -\xi\vec e_\xi\right)\>.
\end{aligned}
\end{equation}
Let $\eta_s(\xi)-\eta=0$ describe the interface, then the normal
vector $\vec n$ can be derived from Frenet's formulas. The position
vector at the interface may be written
\begin{align}
 \vec x=\eta_s \xi\vec e_x+\frac 12\lr{\eta_s^2 -\xi^2}\vec e_y
\end{align}
and the differential line element along this curve is
\begin{align}
 \D s=\sqrt{\D x^2+\D y^2}=\sqrt{\lr{\eta_s^2+\xi^2}\lr{1+\eta_s'^2}}\,\D\xi.
\end{align}
The tangential vector at the interface is given by $\D\vec x/\D s$,
the normal vector must be orthogonal to it. By this condition, it is  determined up to
a sign that we choose so as to make the normal vector point into the
liquid. This procedure yields
\begin{align}
 \vec n=\frac{1}{\sqrt{1+\eta_s'^2}}\left(\vec e_\eta-\eta_s'\vec e_\xi\right)\>.
\end{align}
 The curvature is given by
\begin{align}
 \kappa&=-\vec n\cdot\abl[2]{\vec r}{s}
\nonumber\\
&=-\frac{1}{\sqrt{\xi^2+\eta_s^2}}\lre{\frac{\eta_s''}{\lr{1+\eta_s'^2}^{\frac{3}{2}}}
+\frac{\eta_s'\xi-\eta_s}{\lr{\xi^2+\eta_s^2}\sqrt{1+\eta_s'^2}}}
\label{eq:curvature}
\end{align}
and it is positive for a convex solid.

 We assume the usual model of four-fold crystalline anisotropy:
\begin{align}
 a(\theta)=1-\beta\cos 4\theta=1-\beta\lr{1-8\cos^2\theta\sin^2\theta}
\end{align}
The small parameter $\beta$ is the strength of the anisotropy.
$\theta$ is the angle of the interface normal with the $y$ axis, so we
have $\cos\theta=\vec n\cdot\vec e_y$ and $\sin\theta=\vec n\cdot\vec
e_x$, which allows us to find the anisotropy function expressed in
parabolic coordinates.
\begin{equation}
 a(\theta)=1-\beta\lre{1-8\frac{\lr{\xi-\eta_s\eta_s'}^2\lr{\eta_s+\xi\eta_s'}^2}{\lr{\xi^2
+\eta_s^2}^2\lr{1+\eta_s'^2}^2}}\label{anisotropy}
\end{equation}

Finally, the flow velocity is given by
\begin{align}
 \vec w=\frac{1}{\sqrt{\xi^2+\eta^2}}\left(\psi_\eta\vec e_\xi-\psi_\xi\vec e_\eta\right)\>.
\end{align}

\section{Derivation of the interface equation}
\label{sec:deriv_interface_eq}

We first introduce some simplifications of notation. Substituting
$\bar\tau=\xi$ and $\bar s'=\frac12(u-\xi)$ and defining
\begin{align}
 F(\xi)&=\sigma\kappa(\xi) a(\theta(\xi)) -\lr{1-\I\xi}h(\xi)\>,\\
 z(\xi)&=1+\I h'(\xi)\>,\\
 \bar z(\xi)&=1-\I h'(\xi)\>,
\end{align}
we have from Eq.~\eqref{eq:Nsol} with \eqref{eq:limN_sbartoinf}
\begin{align}
  F'(\xi)&=-\frac{\I}{2}P_f\xi\bar z(\xi)\int\limits_\xi^{\I\infty}M\lr{\frac{1}{2}(\xi-u),u}\D u
\nonumber\\
&-\frac{1}{2}P_f\bar z(\xi)\int\limits_\xi^{\I\infty}\frac{\lre{uh}'}{z(u)}
\EXP{\frac{P_f}{8}\lr{\xi -u}^2}\D u
\label{eq:F1}
\end{align}
with the prime denoting a derivative with respect to $\xi$ or $u$,
depending on whether the term concerned is outside or inside an integral.
 Writing out $M$, we have 
\begin{align}
  &M\lr{\frac{1}{2}(\xi-u),u}=\frac{\I}{2z(u)}\EXP{\frac{P_f}{8}\lr{\xi^2-u^2}}
  \nonumber\\
  &\hspace{5mm}\times\lre{\lre{\lr{1+\I u}h}'-
    \frac{2\lre{uh}'}{u}\lr{1-\EXP{-\frac{P_f}{4}\lr{\xi-u}u}}}
\end{align}
and
\begin{align}
 \pabl{}{\xi}M\lr{\frac{1}{2}(\xi-u),u}&=\frac{P_f\xi}{4}M\lr{\frac{1}{2}(\xi-u),u}
  \nonumber\\
&\;\;-\frac{\I P_f}{4}\frac{\lre{uh}'}{z(u)}\EXP{\frac{P_f}{8}\lr{\xi-u}^2}.
\end{align}
Inserting this into (\ref{eq:F1}), we obtain a useful expression
for the derivative of $F$:
\begin{align}
  F'(\xi)&=-2\I\bar z(\xi)\int\limits_\xi^{\I\infty}\pabl{}{\xi}M\lr{\frac{1}{2}(\xi-u),u}\D u
\nonumber\\
&=-2\I\bar z(\xi)\lre{\pabl{}{\xi}\int\limits_\xi^{\I\infty}M\lr{\frac{1}{2}(\xi-u),u}\D u
+M(0,\xi)}
\nonumber\\
&=-2\I\bar z(\xi)\pabl{}{\xi}\int\limits_\xi^{\I\infty}M\lr{\frac{1}{2}(\xi-u),u}\D u
\nonumber\\
&\hspace{5mm}+\frac{\bar z(\xi)}{z(\xi)}\lre{\lr{1+\I\xi}h(\xi)}'.
\end{align}
This can be integrated by parts. Using
\begin{align}
 \lre{\frac{\bar z}{z}}'=-2\I\frac{h''}{z^2}\>,
\end{align}
we arrive at
\begin{align}
  F(\xi)=&-2\I\bar z(\xi)\int\limits_\xi^{\I\infty}M\lr{\frac{1}{2}(\xi-u),u}\D u
\nonumber\\
&+2\int\limits^\xi h''(\xi')\int\limits_{\xi'}^{\I\infty}M\lr{\frac{1}{2}(\xi'-u),u}\D u\,\D\xi'
\nonumber\\
&+\frac{\bar z(\xi)}{z(\xi)}\lr{1+\I\xi}h(\xi)
\nonumber\\
&+2\I\int\limits^\xi \frac{h''(\xi')}{z^2(\xi')}\lr{1+\I\xi'}h(\xi')\D\xi'\>,
\label{eq:F2}
\end{align}
which is not quite the form we want. On the one hand, equation
(\ref{eq:F2}) manifests a certain generality, since it is valid for
  arbitrary flows. But on the other hand,
we would appreciate to have a right hand side that obviously vanishes for $P_f\to
0$. To achieve this, we use (\ref{eq:F2}) to eliminate the first term on
the right hand side of (\ref{eq:F1}):
\begin{align}
F'(\xi)=&\frac{P_f}{4}\xi F(\xi)-\frac{P_f}{4}\xi\frac{\bar z(\xi)}{z(\xi)}\lr{1+\I\xi}h(\xi)
\nonumber\\
&-\frac{\I}{2}P_f\xi\int\limits^{\xi}\frac{h''(\xi')}{z^2(\xi')}\lr{1+\I\xi'}h(\xi')\D\xi'
\nonumber\\
&-\frac{P_f}{2}\xi\int\limits^{\xi}h''(\xi')\int\limits_{\xi'}^{\I\infty}M\lr{\frac{1}{2}(\xi'-u),u}\D u\,\D\xi'
\nonumber\\
&-\frac{P_f}{2}\bar z(\xi)\int\limits_\xi^{\I\infty}\frac{\lre{uh}'}{z(u)}\EXP{\frac{P_f}{8}\lr{\xi -u}^2}\D u
\label{eq:diffeqF}
\end{align}
Employing the identities
\begin{align}
 \int\limits_\xi^{\I\infty}\frac{\lre{uh}'}{z(u)}\EXP{\frac{P_f}{8}\lr{\xi -u}^2}\D u&=-\frac{\xi h(\xi)}{z(\xi)}
\nonumber\\
&\hspace*{-25mm}-\int\limits_\xi^{\I\infty}\lre{uh(u)\lr{\frac{\lr{u-\xi}P_f}{4z(u)}-\frac{\I h''(u)}{z^2(u)}}\EXP{\frac{P_f}{8}\lr{\xi -u}^2}}\D u\>,
\\
F'(\xi)-\frac{P_f}{4}\xi F(\xi)&=\EXP{\frac{P_f}{8}\xi^2}\abl{}{\xi}\lre{F(\xi)\EXP{-\frac{P_f}{8}\xi^2}}\>,
\end{align}
the  first of which is obtained via integration by parts again, 
we may rewrite \eqref{eq:diffeqF} as follows
\vspace*{8mm}\\
\vspace*{-6mm}
\begin{align}
  \EXP{\frac{P_f}{8}\xi^2}&\abl{}{\xi}\lre{F(\xi)\EXP{-\frac{P_f}{8}\xi^2}}=\,\frac{P_f}{4}\frac{\bar
    z(\xi)}{z(\xi)}\xi\lr{1-\I\xi}h(\xi)
  \nonumber\\
  &+\frac{P_f}{2}\bar
  z(\xi)\int\limits_\xi^{\I\infty}\left[\rule{0pt}{6mm}\right.uh(u)\left(\rule{0pt}{4.5mm}\right.
  \frac{\lr{u-\xi}P_f}{4z(u)}
  \nonumber\\
  &\hspace*{20mm}-\frac{\I
    h''(u)}{z^2(u)}\left.\rule{0pt}{4.4mm}\right)\EXP{\frac{P_f}{8}\lr{\xi
      -u}^2}\left.\rule{0pt}{6mm}\right]\D u
  \nonumber\\
  &-\frac{\I}{2}P_f\xi\int\limits^{\xi}\frac{h''(\xi')}{z^2(\xi')}\lr{1+\I\xi'}h(\xi')\D\xi'
  \nonumber\\
  &-\frac{P_f}{2}\xi\int\limits^{\xi}h''(\xi')\int\limits_{\xi'}^{\I\infty}M\lr{\frac{1}{2}(\xi'-u),u}\D
  u\,\D\xi'.
\end{align}
\vspace*{6mm}
With one further integration, we arrive at Eq.~\eqref{eq:F3}.
\\
\end{appendix}


%

\end{document}